\documentclass[preprint2]{aastex}

\shorttitle{Photometry and Redshifts in the Lockman Hole}
\shortauthors{Fotopoulou et al.}
\begin{document}
\title{Photometry and Photometric Redshift catalogs for the Lockman Hole Deep Field}
\author{S. Fotopoulou\altaffilmark{1,2},
M. Salvato\altaffilmark{1},
G. Hasinger\altaffilmark{1,3},
E. Rovilos\altaffilmark{4,5},
M. Brusa\altaffilmark{4},
E. Egami\altaffilmark{6},
D. Lutz\altaffilmark{4},
V. Burwitz\altaffilmark{4},
J.H. Huang\altaffilmark{7},
D. Rigopoulou\altaffilmark{8,9},
M. Vaccari\altaffilmark{10}}

\altaffiltext{1}{Max Planck Institut f\"ur Plasma Physik Boltzmannstrasse 2, 85748 Garching Germany}
\altaffiltext{2}{Technische Universit\"at M\"unchen, James-Franck-Strasse 1, 85748 Garching Germany}
\altaffiltext{3}{Institute for Astronomy, University of Hawai, Honolulu, USA}
\altaffiltext{4}{Max Planck Institut f\"ur Extraterrestrische Physik Giessenbachstrasse, 85748 Garching Germany}
\altaffiltext{5}{INAF-Osservatorio Astronomico di Bologna, Via Ranzani 1, 40127, Bologna, Italy}
\altaffiltext{6}{Steward Observatory, University of Arizona, 933 North Cherry Avenue, Tuscon, AZ 85721, USA}
\altaffiltext{7}{Harvard-Smithsonian Center for Astrophysics, 60 Garden Street, Cambridge, MA 02138, USA}
\altaffiltext{8}{Space Science \& Technology Department, Rutherford Appleton Laboratory, Chilton, Didcot, Oxfordshire OX11 0QX, UK}
\altaffiltext{9}{Astrophysics, Oxford University, Keble Road, Oxford OX1 3RH, UK}
\altaffiltext{10}{Dipartimento di Astronomia, Universit\`a di Padova, vicolo Osservatorio, 3, 35122 Padova, Italy}

\email{sotiria.fotopoulou@ipp.mpg.de}

\begin{abstract}
We present broad band  photometry and photometric redshifts for 187611
sources  located in  $\sim0.5\,\rm{deg}^2$ in  the Lockman  Hole area.
The catalog  includes 389 X-ray  detected sources identified  with the
very   deep  XMM-Newton   observations  available   for  an   area  of
$0.2\,\rm{deg}^2$.  The  source detection was performed  on the $R_c$,
$z'$ and $B$ band images and the available photometry is spanning from
the far  ultraviolet to  the mid infrared,  reaching in the  best case
scenario 21 bands. Astrometry corrections and photometric cross-calibrations
over the entire dataset allowed  the computation of accurate photometric redshifts.
Special treatment is undertaken  for the X-ray sources, the  majority
of which is active galactic nuclei.   Comparing the photometric redshifts
to  the available  spectroscopic  redshifts we  achieve for  normal
galaxies  an  accuracy  of  $\sigma_{\Delta z  /(1+z)}=0.036$,  with
$12.7\%$ outliers, while for the X-ray detected sources the accuracy
is $\sigma_{\Delta  z /(1+z)}=0.069$, with  $18.3\%$ outliers, where
the      outliers     are      defined      as     sources      with
${|z_{phot}-z_{spec}|}>0.15\cdot{(1+z_{spec})}$.  These results are
a  significant improvement over  the previously  available photometric
redshifts for  normal galaxies  in the Lockman  Hole, while it  is the
first time that photometric redshifts are computed and made public for
AGN for this field. 

\end{abstract}
\keywords{galaxies:active - galaxies:general - galaxies:photometry - surveys}

\section{Introduction}
\label{sec:introduction}
Multiwavelength  surveys  provide  the most  successful  observational
strategy towards  the understanding of  galaxies. The multiwavelength
coverage provides the opportunity to compute photometric redshifts and
to study the Spectral Energy Distributions (SED) for a large number of
galaxies. At the same time, careful estimation of the distances of the
galaxies enables the study of  the intrinsic properties of the sources
such  as luminosities  and  the determination  of  the evolution  with
redshift of other fundamental  parameters such as stellar masses, star
formation  rates, etc. Furthermore,  distance measurements  enable the
use  of galaxies  as cosmological  probes, for  example  through their
clustering properties.

The distance measurement for galaxies is based on the determination of
the redshift of the source, which is specified very accurately through
spectroscopy.  However, for large samples of sources either in deep
  pencil beam fields, or shallower but more extended fields, the most
  efficient way  to compute the distance is  via photometric redshifts
  \citep[e.g.][]{Budavari00}, although  their
  accuracy  is strongly  depending on  i) the  number and  the type  of
  filters  (broad-band versus  intermediate-band),  ii) the  redshift
  range and type  of galaxies of interest (passive, versus starforming  
  or active galactic nuclei (AGN)).  

For  fields where  extensive  photometric datasets  are available  the
photometric  redshift  technique   has  been  employed  with  reliable
results.   For   example   in   CFHTLS\footnote{Canadian-France-Hawaii
  Telescope        Legacy        Survey}       \citep{Il06},        in
AEGIS\footnote{All-wavelength   Extended  Groth   strip  International
  Survey} \citep{Barro11}, in COSMOS\footnote{Cosmic Evolution Survey}
\citep{Il09,  Sal09}, in FDF\footnote{FORS Deep Field} \citep{Bender01}, 
in the  CDFN\footnote{Chandra Deep  Field North}
\citep{Barger03}  and in the  CDFS\footnote{Chandra Deep  Field South}
\citep{Wolf04, Luo10, Car10} the  photometric redshifts  reached  an accuracy
$\Delta  (z_{\rm   phot}-z_{\rm  spec})/(1+z_{\rm  spec})<0.03$,  thus
allowing (among other  applications) a 3D mapping of  the Dark Matter
\citep{Massey07}, studies  of the evolution of luminosity functions of normal galaxies
\citep[e.g.][]{Ilbert06, Gabasch04, Gabasch06, Caputi06}  and  AGN  
\citep[e.g.][]{Hasinger05, Barger05, Ebrero09, Aird10}, the first determination of 
the high-z (z$>$3) logN-logS and space density from X-ray selected AGN 
\citep{Brusa09, Civano11}, Compton thick objects \citep[e.g.][]{Fiore09,Luo11}, etc. 
Moreover,  photometric redshifts are  also used for the  study of
groups and clusters of galaxies \citep{Giodini09,Papovich10,Geach11} 
and much more.

Yet, not  all  deep and wide fields have photometric  redshifts of
comparable accuracy,  thus limiting any  study to the  parameter space
allowed by a given field. Only  reliable photometric redshifts in all 
the fields will allow a complete exploitation of  the total telescope 
time devoted to survey studies in the last ten  years.  Keeping this 
is mind, here we present optically  based   extensive  photometry  and
photometric redshift catalogs for the Lockman Hole Deep Field.

The  Lockman Hole  is the  area on  the sky  with the  lowest galactic
hydrogen    column     density    along    the     line    of    sight
($N_H\approx5.7\,10^{19}\rm{cm}^{-2}$,    \citealt{Lock86,Schlegel98}).
This  physical  characteristic  provides  the opportunity  to  perform
extragalactic  observations  without  significant  absorption  of  the
radiation in  the soft  X-rays and the  ultra-violet and  with minimal
galactic cirrus emission  in the infrared. For this  reason the field
has been observed in all energy bands from the Radio
to  the X-rays.  A detailed  overview of  the various  observations is
given in \citet{Rov09}.

Up  to  now the  spectroscopic  follow  up  was dedicated  to  sources
detected in specific bands  (X-ray, infrared, radio). In addition, the
photometry of  the field was  lacking crucial bands  such as J  and K,
hampering the  possibility to compute  accurate photometric redshifts.
The only  public photometric redshifts  for approximately half  of the
area of  the field examined  in this work  are available via  the SWIRE
survey  \citep{Row08},  computed  using  photometry in  4  broad  band
optical filters ({\it U', g', r', i'})  and 3.6$\mu m$ and 4.5$\mu m$
mid-infrared  filters   from  {\it  Spitzer/IRAC}.   Furthermore,  the
photometric  redshifts were  lacking proper  treatment for  the AGN.

In  this work,  we  first cross-calibrate  and  then combine  publicly
available  and private  photometry from  the ultra-violet  to  the mid
infrared wavelengths, creating an  extensive catalog, which we further
use  to compute photometric  redshifts.  We  also separate  the sample
according  to the  X-ray emission  and use  the proper  combination of
templates and priors to achieve the best photometric redshift solution
for all sources. In addition to photometry and photometric redshifts we
release an  up-to-date list of the spectroscopic  data, both public and
private, available in the field.

The  layout  of the  paper  is as  follows:  in  \S \ref{sec:data}  we
describe the  data we used for  this work. In  \S \ref{sec:catalog} we
describe the processing of the images and the procedure we followed in
order to  create our catalog.  In \S \ref{sec:photoz} we  describe the
method  we   used  to  compute   the  photometric  redshifts.   In  \S
\ref{sec:results} we present the photometric redshifts and discuss our
results   comparing  also   with  previous   works.  Finally,   in  \S
\ref{sec:conclusions} we  present our conclusions.  All magnitudes are
expressed in the AB system \citep{Oke83}.

\placetable{tab:fields}

\section{Data Set}
\label{sec:data}
\begin{figure}
\epsscale{0.80}
\includegraphics[width=\linewidth]{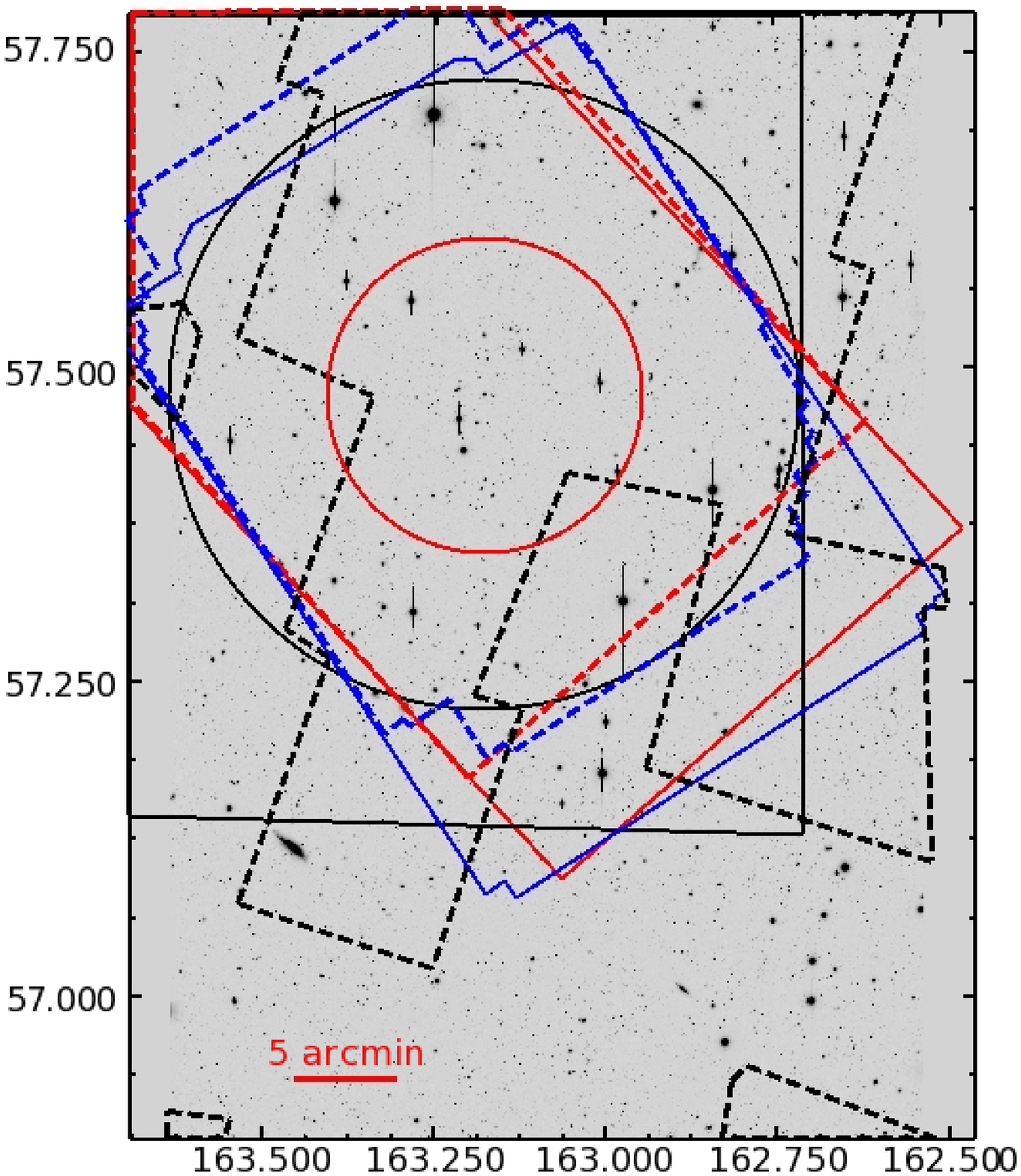}
\caption{The area coverage  of the Lockman Hole Deep  Field. The image
  denotes the  area observed  in the $R_c$,  $I_c$, $z'$  filters with
  Subaru.  The marked  regions represent:  U and  B LBT  filters (blue
  solid line), V, Y and z' LBT filters (blue dashed line), $\rm{3.6\mu
    m}$ and $\rm{5.8\mu m}$ IRAC (red solid line), $\rm{4.5\mu m}$ and
  $\rm{8\mu m}$ IRAC  (red dashed line), J and  K UKIRT filters (black
  solid  line). Even though the whole area is covered by SDSS, only the 
  area enclosed in the black dashed outline has good photometry 
  (flag=3 in the SDSS catalog). The black  circle  represents  the  area targeted  by
  XMM-Newton and  the red circle represents roughly  the area targeted
  by the Hubble Space Telescope.
\label{fig:area}}
\end{figure}

\begin{figure}
\includegraphics[width=\linewidth]{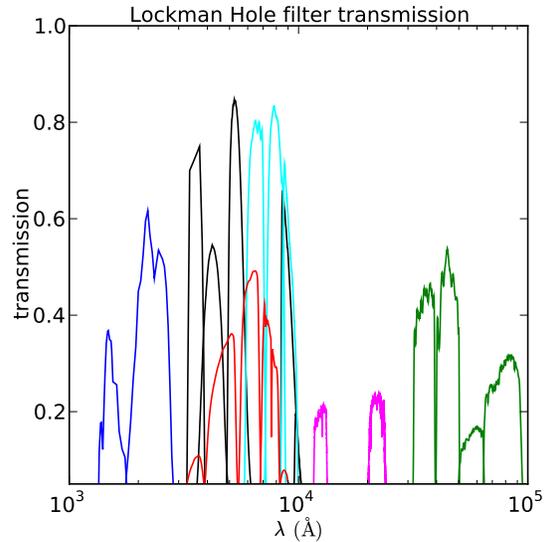}
\caption{The filter coverage of the Lockman Hole Deep Field. Blue: GALEX (FUV, NUV), black: LBT (U, B, V, Y, z'), cyan: Subaru($\rm{R_c}$, $\rm{I_c}$, $\rm{z'}$), red: SDSS ($\rm{u'}$, $\rm{g'}$, $\rm{r'}$, $\rm{i'}$, $\rm{z'}$), magenta: UKIDSS (J, K), green: Spitzer ($\rm{3.6\mu m}$, $\rm{4.5\mu m}$, $\rm{5.8\mu m}$, $\rm{8.0\mu m}$).
\label{fig:filters}}
\end{figure}

In this section we describe the data that we used in each energy band,
separated by  instrument/satellite. In Fig. \ref{fig:area}  we give an
overview of the coverage of Lockman  Hole area in each every band, while in
Fig.   \ref{fig:filters}   we    present   the   filter   transmission
curves.

\placefigure{fig:area}
\placefigure{fig:filters}

\subsection{X-ray data}
The field has been observed with XMM-Newton in the time period between
April 2000 and December 2002, with a total raw exposure of 1.30 Ms for
the detectors  on board  the satellite (EPIC  MOS and EPIC  pn).  The
field is centered at $\rm{+10^h52^m43^s,\,+57^{\circ}28'48''}$ and has
a  radius of $15'$,  thus covering  an area  of $\rm{\sim0.2\,deg^2}$.
The     limiting     flux     in     each    detection     band     is
$\rm{F_{0.5-2.0\,keV}=1.9\cdot10^{-16}erg\,cm^{-2}s^{-1}}$ in the soft
band,   $\rm{F_{2.0-10.0\,keV}=9\cdot10^{-16}erg\,cm^{-2}s^{-1}}$,  in
the                   hard                   band                  and
$\rm{F_{5.0-10.0\,keV}=1.8\cdot10^{-15}erg\,cm^{-2}s^{-1}}$,   in  the
ultra hard band.   The depth of the X-ray  observations in combination
with the size  of the area place the field close  to the Extended CDFS
(ECDFS),  between the  very deep  pencil-beam fields  (e.g.   CDFN and
CDFS) and  the more extended but shallower  fields (e.g.  XMM-COSMOS).
In particular, the X-ray observations  in the Lockman Hole Field reach
one  order of  magnitude fainter  flux  limit than  XMM-COSMOS in  all
detection bands  (Table \ref{tab:fields}).   The catalog of  the X-ray
detected sources is presented in  \citet{Br08} and contains a total of
409 sources  with likelihood of  being real detections greater  than 10
(3.9$\sigma$).   The  catalog of  optical  counterparts  of the  X-ray
sources is  presented in  \citet{Rov11} (see also  \S \ref{sec:Xcat}).
In summary, the counterparts were assigned on the basis of a Likelihood 
Ratio (LR) technique  applied   to  our   optical  and
near-infrared catalogs,  which reach a depth of  $\rm{R_{c,lim}}$ = 26
mag and [3.6$\mu m$]$_{lim}$ = 24.6 mag.

\subsection{Ultra-violet data}
In  the far  ultra-violet (FUV)  and the  near ultra-violet  (NUV) the
Lockman  Hole  area was  observed  by  the  Galaxy Evolution  Explorer
(GALEX).   Here,  we consider  the  magnitude  in $\rm{3''}$  diameter
aperture     available     in      the     General     Release     4/5
(GR4/5)\footnote{http://galex.stsci.edu/GR4/}.     To    correct   the
aperture   photometry   to   total,   we  followed   the   recipe   of
\citet{Morr07}. The authors presented  the growth curve analysis using
as targets white dwarfs and we use the same correction factors through
out the  whole catalog.  The Lockman  Hole is part  of the  GALEX Deep
Imaging Survey  and the third  quartile of the  magnitude distribution
reaches $\rm{FUV_{lim}=24.5\,mag}$ and $\rm{NUV_{lim}=24.5\,mag}$.

\subsection{Optical data}
\begin{figure}
\includegraphics[width=\linewidth]{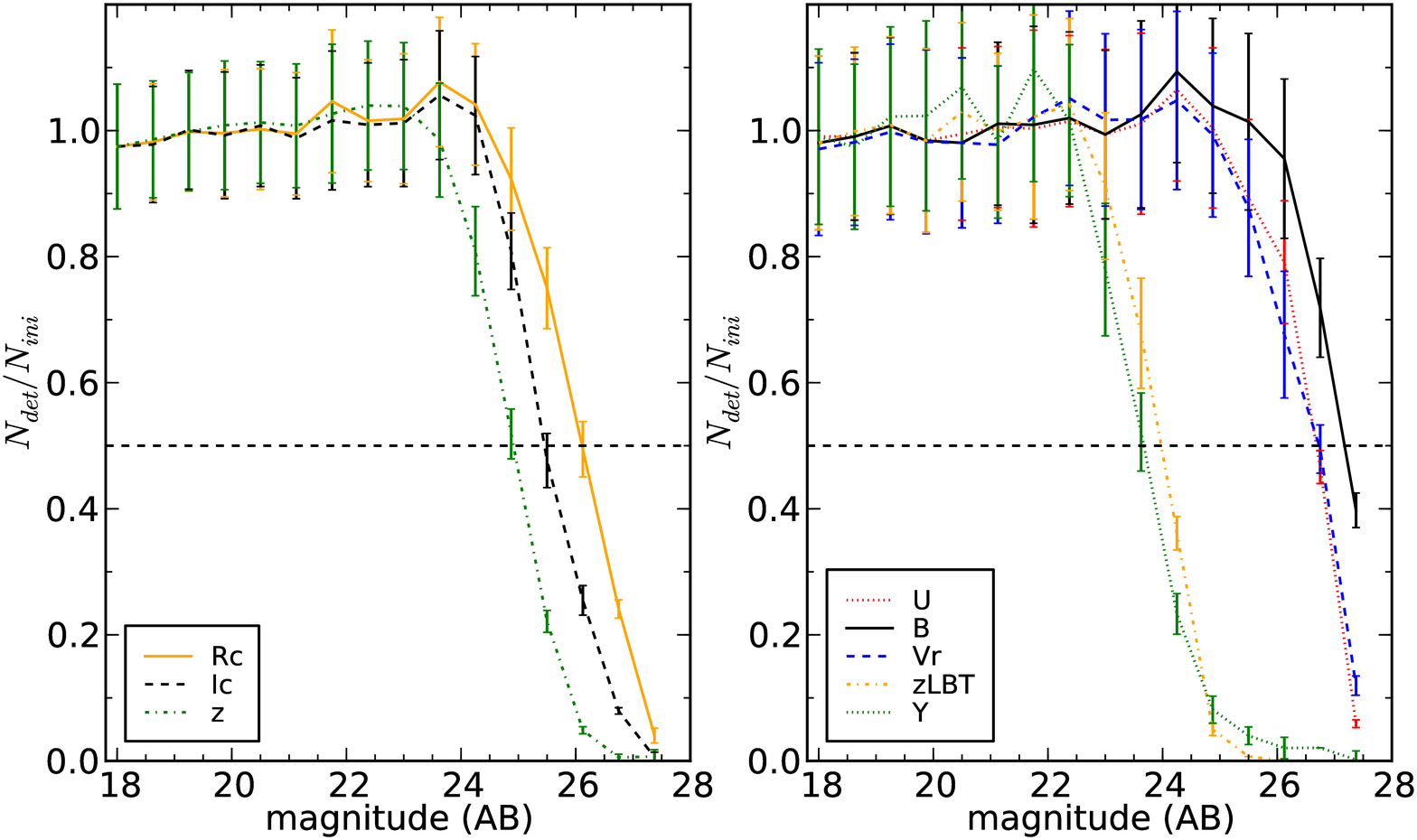}
\caption{The completeness curves for the Subaru (left panel) and LBT (right panel) images, computed from simulated sources. The horizontal dashed line marks the 50\% completeness limit.
\label{fig:completeness}}
\end{figure}

In the optical  wavelengths, our dataset consists of  a compilation of
observations from LBT,  Subaru and SDSS. Our images  exhibit very good
seeing\footnote{Calculated from the mean PSF FWHM of 30 stars.} of the
order of $\rm{0.9''-1''}$, such that we could retrieve almost
$100\%$ of the flux  for point-like sources within $\rm{3''}$ aperture
(see  \S  \ref{sec:OptCat}  for  details).  Thus,  we  are  using  the
$\rm{3''}$  aperture magnitudes  without applying  any  corrections to
total. In Fig. \ref{fig:completeness}, we present the completeness analysis
for all  optical filters.  We plot the  ratio of detected  over true
  number of simulated point-like sources, versus magnitude.
 Details on the area  observed, the total exposure time, the PSF
full width half maximum (FWHM), the 50\% detection magnitude limit for
point-like sources ($\rm{5\sigma}$, AB)  and the AB to Vega correction
factor for each filter,  are presented in Table \ref{tab:LimMag}. 

\placefigure{fig:completeness}
\notetoeditor{figure fig:completeness should span two columns.}

\subsubsection{Large Binocular Telescope}
We observed the  Lockman Hole in the period  between February 2007 and
March  2009  in  5  bands  ($\rm{U}$,  $\rm{B}$,  $\rm{V}$,  $\rm{Y}$,
$\rm{z'}$) using the Large  Binocular Telescope (LBT) covering an area
of about $0.25\,\rm{deg^2}$.  The reduction of the data and the number
counts of  the very deep observations in  the U, B and  V filters have
been  published in \citet{Rov09}.   The two  telescopes of  LBT have
slightly  different cameras,  one optimized  for bluer  and  the other
optimized for redder bands. In  \citet{Rov09} the published V band is 
the combination of two separate images ($\rm{V_{blue}}$, $\rm{V_{red}}$).
However, here we are using only  the $\rm{V_{red}}$  image, as the 
differences in  the two filter curves and the low quality of the 
$\rm{V_{blue}}$  image,  make the  stacking  not an option.

In addition to U, B and V photometry, in this work we also include the
shallower  observations in  the Y  and z'  filters which  were reduced
later in a similar manner.

\subsubsection{Subaru}
Complementary to  our own  photometry,  we made  use of  data from  the
Institute for Astronomy (Hawaii) Deep Survey. Observations with the Subaru telescope
between November 2001  and April 2002 provided imaging  in an area of
$0.53\,\rm{deg^2}$. The  observations were carried  out in the  Rc, Ic
and z' filters.  Details about the survey and  the observations on the
field can be found in \citet{Bar04}, while details on the analysis of 
the data can be found in \S \ref{sec:OptCat}.

\subsubsection{SDSS}

Unfortunately SDSS  does not cover the Lockman Hole field with uniform 
photometric quality. However,  we include SDSS  (DR7, \citealt{Ab09})
$\rm{u'}$, $\rm{g'}$, $\rm{r'}$,  $\rm{i'}$, $\rm{z'}$ photometry 
(fiber magnitude) when available, if $\rm{r_{AB}'<22\,mag}$ and 
photometric quality flag = 3, provided in the SDSS catalog 
(Fig. \ref{fig:area} black dashed line). The  first
reason to use  the SDSS catalog is that the  observations from LBT and
Subaru     are    very    deep     and    the     brightest    sources
($\rm{R_c\approx18\,mag}$) are saturated. Thus the shallower SDSS data
can provide  a solution to  saturation problems.  For example,  in the
$\rm{R_c}$ filter, 677 sources are flagged as saturated and photometry
is not available.  However, for 170 of these sources SDSS photometry is
available and can be used to  recover the SEDs and thus the photometric
redshifts.

SDSS photometry is especially important for X-ray detected sources, for
which we  want to trace optical variability.   This intrinsic property
of AGN  dominated systems, can  affect the computation  of photometric
redshifts,  when  the photometry  is  obtained non-simultaneously  and
through  multi-epoch  observations.  As   we  explain  further  in  \S
\ref{sec:Xcat},  we use the  $\rm{z'}$ band  to detect  the variability  of a
source,  since  in  this   filter  we  have  observations  from  three
telescopes at different epochs.

\subsection{Near and Mid infrared data}
\subsubsection{UKIDSS data}
The near-infrared view of our  sources is provided by the data release
7 of the UKIRT Infrared Deep Sky Survey. The UKIDSS project is defined
in  \citet{Law07}. UKIDSS  uses the  UKIRT Wide  Field  Camera (WFCAM;
\citealt{Cas07}).    The   photometric    system   is   described   in
\citet{Hew06}  and   the  photometric  calibration   is  explained  in
\citet{Hodgkin09}.  The Lockman Hole is part of the Deep Extragalactic
Survey to be observed in the J, H  and K bands.  Up to now, only the J
and K  bands are available with limiting  magnitudes of $J_{lim}=23.4$
mag   and   $K_{lim}=22.9$    mag   (5$\rm{\sigma}$,   point   source)
respectively.  We use the  $\rm{2.8''}$ aperture magnitude provided in
the  catalog  which  is  already  corrected to  total  for  point-like
sources. The aperture corrections were determined from the growth curve
analysis of bright stars as described in \citet{Hodgkin09}.

\subsubsection{Spitzer data}
For the near to mid  infrared wavelengths we are using observations at
$\rm{3.6\mu  m}$, $\rm{4.5\mu  m}$, $\rm{5.8\mu  m}$,  $\rm{8.0\mu m}$
from  the Infrared  Array Camera  (IRAC). Although the
IRAC images are the  same as the images used in  \citet{Per08},  we 
created our own catalog.  As explained in \citet{Rov11}
we extracted the  IRAC photometry using SExtractor in  dual mode using
the $\rm{3.6\mu m}$ image to detect the sources in  all other IRAC filters.
Following \citet{Sur05},  we used the  $\rm{2.8''}$ diameter aperture magnitude
and   applied   the  same   corrections   calculated  for   point-like
sources.  The  aperture corrections  have  been determined  performing
growth curve analysis of a composite PSF consisting of 10-20 stars.

The  50\% efficiency limiting  magnitude of  the IRAC  observations is
$24.6$ mag for the $\rm{3.6\mu  m}$ band (2$\sigma$).  We verified our
photometry      against     the      publicly      available     SWIRE
catalog\footnote{http://irsa.ipac.caltech.edu/data/SPITZER/SWIRE/}   in
the  Lockman  Hole,  which  is  shallower  and  complementary  to  our
observations.   Using approximately  2400 sources  in  the overlapping
$\rm{0.25deg^2}$ we find that the orthogonal distance regression gives
slope  of  $0.995\pm0.002$,  intercept $0.19\pm0.05$  and  correlation
length $0.73$, in agreement with  what is found by \citet{Per08}.  The
offset between our photometry  and the SWIRE photometry is explainable
in view  of the different,  continuously improved pipeline used  by the
IRAC team in reducing the data.

\placetable{tab:LimMag}

\section{Catalog Assembly}\label{sec:catalog}

The    photometric  catalog    that   we    release    in   electronic
version\footnote{but   which   will   be   periodically   updated   at
  http://www.rzg.mpg.de/$\sim$sotiriaf/surveys/LH/},  covers  the area
observed by Subaru (Fig. \ref{fig:area})  which is the widest in terms
of area ($\sim0.5\,\rm{deg}^2$).  Only 42\% of the Subaru observations
are  covered by  the XMM-Newton  observations, and  for  the remaining
field we do not have information  on the presence of an AGN.  Thus, we
decided to flag the sources outside the XMM-Newton area.

In  this section  we describe  the re-processing  and  the photometric
reduction  of the LBT  and Subaru  images and  the compilation  of the
catalogs.

\subsection{Image re-processing}

We  re-processed  the  LBT  ($\rm{U}$, $\rm{B}$,  $\rm{V}$,  $\rm{Y}$,
$\rm{z'}$)  and Subaru ($\rm{R_c}$,  $\rm{I_c}$, $\rm{z'}$)  images to
make them  as uniform as  possible. First, in  order to bring  all the
optical  images to a  common grid,  we registered  the LBT  and Subaru
images to the SWIRE astrometry  using the 'Geomap' and 'Geotran' tasks
in IRAF with a 6th  order polynomial transformation to correct for any
residual distortions.  The final  relative astrometric accuracy  is of
the order of $\rm{0.2''}$, equal to the pixel scale of the images \citep{Rov11}.

Furthermore, in order to  retrieve homogeneous photometry we convolved
all  the  images  to  the  largest  seeing  ($\rm{1.06''}$)  using  an
appropriate  Gaussian kernel  for each  image in  the task  'Gauss' in
IRAF.

\subsection{Catalog compilation}

\begin{figure}
\includegraphics[width=\linewidth]{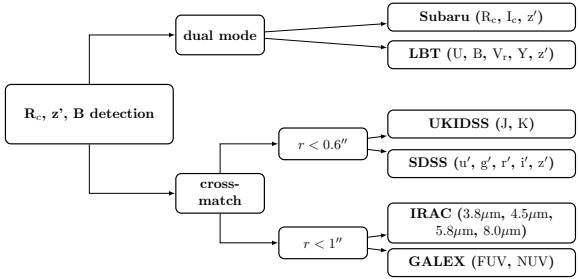}
\caption{Schematic diagram of the merging procedure of the various catalogs. Using as base images the $\rm{R_c}$, $\rm{z'}$, $\rm{B}$ images, we calculate in dual mode the photometry from the LBT and Subaru images. We perform positional matching with the independently obtained photometric catalogs from UKDISS, SDSS, IRAC, and GALEX. The X-ray catalog is then matched to the multi-wavelength optical catalog using Likelihood Ratio matching as described in \citet{Rov11}. A detailed description is given in \S \ref{sec:OptCat}.
\label{fig:cat_merge}}
\end{figure}

In  the next  subsections, we  describe the  procedure we  followed to
obtain  the optical  photometry  and as  well  as the  merging of  the
catalogs at all wavelengths. A schematic of the merging is depicted in
Fig. \ref{fig:cat_merge}.  Additionally, we describe  the subsample of
sources   for  which  spectroscopic   redshifts  are   available.  The
compilation  and assembling  of the  final catalogs  is  performed 
using the  publicly available codes  for data mining  in astrophysics,
STILTS\footnote{http://www.starlink.ac.uk/stilts/}  \citep{STILTS} and
TOPCAT\footnote{http://www.starlink.ac.uk/topcat/} \citep{TOPCAT}.

\placefigure{fig:cat_merge}

\subsubsection{Optical Catalog}\label{sec:OptCat}

For all optical images,  we calculate the $\rm{3''}$ diameter aperture
magnitude  using  SExtractor  \citep{SeX}   in  dual  mode,  using  as
reference  for the   source  detection  the  $\rm{R_c}$,
$\rm{z'}$, $\rm{B}$  images.  The dual mode  approach, guarantees that
the photometry is  measured on every image in  an aperture centered at
the  same pixel  position  as  the detection  image.  To take  into
account blended sources we are using during the  detection
 the 'Mexican Hat' filter, recommended for crowded fields.

\begin{figure}
\includegraphics[width=\linewidth]{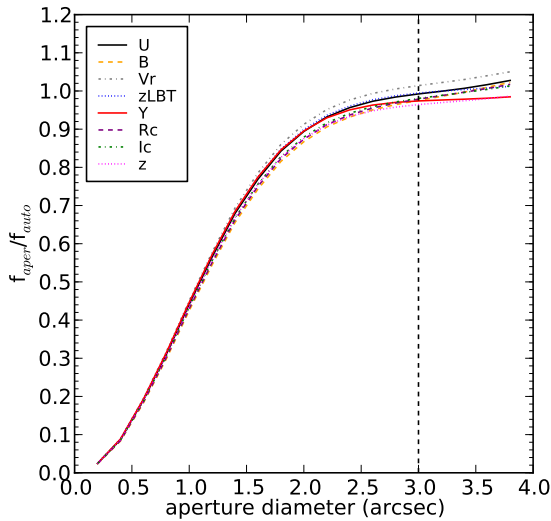}
\caption{Mean growth curves for simulated point sources on the LBT and Subaru images. At least $96\%$ of the total flux is recovered at $3''$ aperture diameter.
\label{fig:growth}}
\end{figure}

In order  to determine  the flux  lost when using  a fixed  aperture (
$\rm{3''}$), we include on  the images, at random positions, simulated
point sources with  a PSF of ${1.06''}$ using  the task 'mkobjects' in
IRAF. In Fig. \ref{fig:growth} we present the growth curve for all the
images. The flux lost is of the order of 0.05 in all cases. This is of
the order of our accuracy and  we choose not to add any corrections in
the optical photometry.

\placefigure{fig:growth}

The final photometric catalog is  obtained by merging the detection on
three  images  and at  two  detection  thresholds, $\rm{5\sigma}$  and
$\rm{3\sigma}$. The detection images are:

\begin{itemize}
\item Priority 1 -- the Subaru Rc image. It is the best image in terms
  of  seeing ($\rm{0.9''}$),  depth  ($\rm{R_{c,lim}=26.1\,mag}$), and
  area  coverage ($\rm{0.53\,deg^2}$). We  retrieve 160633  sources at
  the  $\rm{5\sigma}$  detection  level  and  257352  sources  at  the
  $\rm{3\sigma}$ level.

\item Priority 2 -- the Subaru z' image. There are sources detected in
  the z' band which fall below the detection limit of the Rc band, due
  to  their  intrinsic  SED.  These  objects could  be  high  redshift
  galaxies and/or galaxies that have  large amounts of dust. We choose
  to perform the detection on  the Subaru over the LBT $\rm{z'}$ image
  because it  covers a  larger area,  it is deeper  and it  has better
  seeing. We  retrieve 127362 sources at  the $\rm{5\sigma}$ detection
  level and 250102 sources at the $\rm{3\sigma}$ level.

\item  Priority 3 --  the LBT  B image.  With LBT  the field  has been
  observed with a  sequence of short exposures which  allowed to reach
  the  same depth  of Subaru  but  with less  saturated sources.   The
  number of sources  we retrieve is 68107 at  $\rm{5\sigma}$, while at
  $\rm{3\sigma}$ we retrieve 105658 sources.

\end{itemize}

In order to create the optical  catalogs, we keep the whole Rc catalog
and include from the z' - and B - based catalogs, sources that are not
present  in  the  Rc  catalog   within  $\rm{0.5''}$  from  the  Rc  -
sources.  This  is  done  for the  $\rm{5\sigma}$  and  $\rm{3\sigma}$
detection thresholds separately.

\begin{figure}
\includegraphics[width=\linewidth]{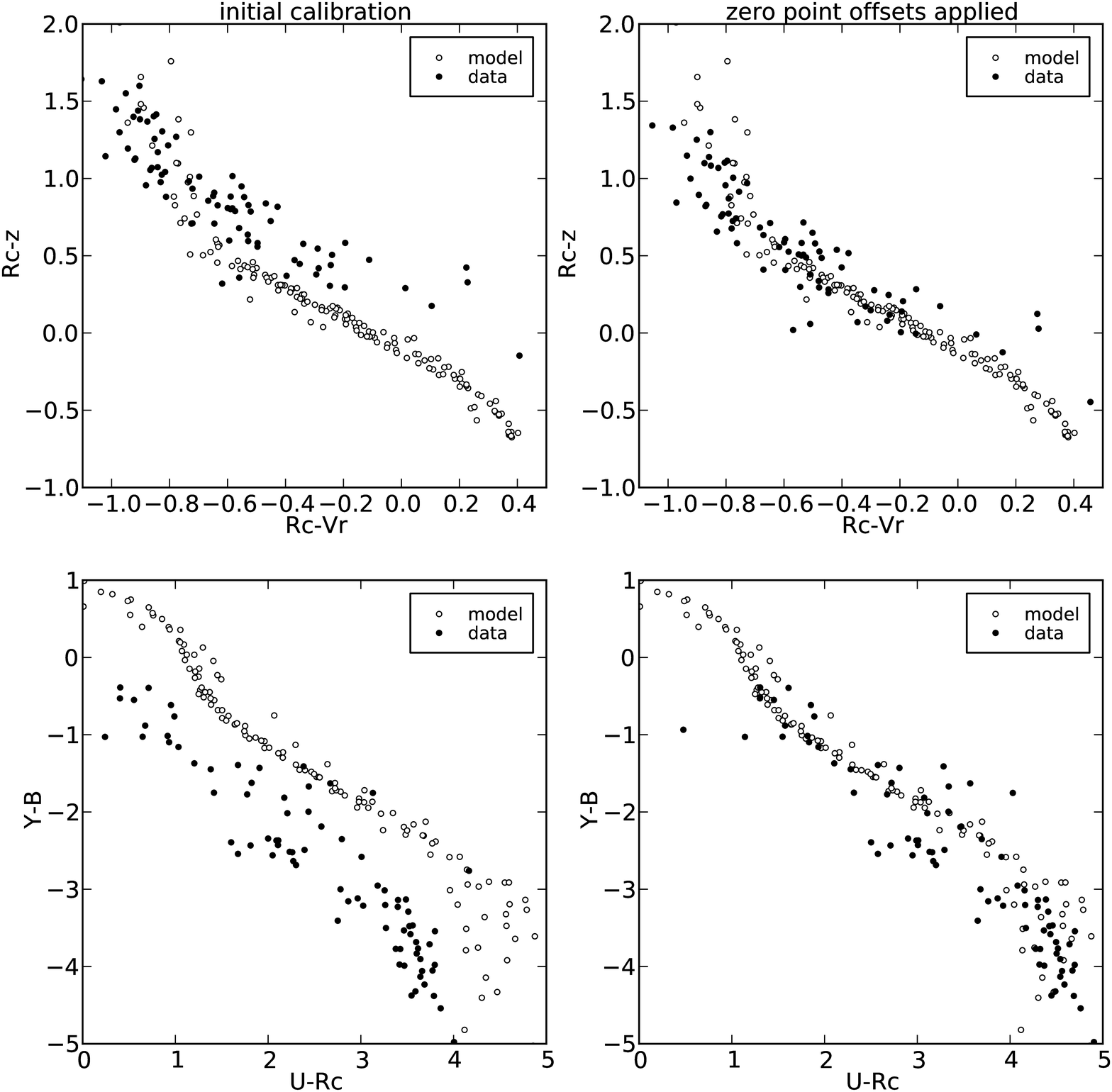}
\caption{We used the star templates of \citet{Pickles98} to check for systematic offsets in the photometric calibration. Two examples of color-color plots of stars for models (open dots) and data (filled dots), before (left panels) and after (right panel) photometric correction are presented here. 
\label{fig:star-tracks}}
\end{figure}

As a  consistency check  of the photometric  calibration and  as cross
calibration   between    different   bands/telescopes,   we   compared
theoretical colors  of stars with  the colors of star  candidates from
our catalog (Fig. \ref{fig:star-tracks}).   In this context, stars are
those   bright  sources   ($\rm{R_c<19}$)   that  SExtractor   defines
point-like ($\rm{R_{class} >0.98}$) by  comparison with the PSF of the
image.  A more refined star/galaxy  separation, on the basis of the SED
fitting will  be performed later on,  in the final  compilation of the
catalog (see section \ref{sec:star}).

After  the  comparison of  the  stellar  photometry  with the  stellar
tracks, a mean correction of the  order of $\sim$0.1 mag is applied to
the   photometry   with   exception   the   Y  band.    As   seen   in
Fig. \ref{fig:star-tracks}, the Y band required a zero point offset of
$\sim0.8$. This was expected, as  the zero point was only approximate,
since no standard stars were observed during the Y band observations.

\placefigure{fig:star-tracks}
\notetoeditor{figure fig:star-tracks should span two columns}

\subsubsection{Combining Optical with Other Wavelengths}

In order  to merge our optical  catalogs with the  GALEX, SDSS, UKIDSS
and IRAC catalogs, we proceed as follows (Fig. \ref{fig:cat_merge}).

We associated all optical sources with the closest GALEX source within
$\rm{1''}$. As  the PSF of  the GALEX images  is much larger  ($\rm{5''}$), a
larger  search radius  should be  adopted.  However,  as  the publicly
available GALEX catalog does not  account for blended sources, a small
radius  reduces  the  number  of false  matches  \citep[see  also][for
  optical to GALEX  associations]{Arnouts05}.  We retrieved $\sim3700$
and $\sim10000$ couterparts in FUV and NUV, respectively.
 
For   the  Sloan   Catalog,  the   matching  radius   is   reduced  to
$\rm{0.6''}$. This corresponds  to 3 pixels on our  optical images. We
keep only  the sources for  which $\rm{r'<22\,mag}$, similarly  to the
selection  criterion  of  \citet{Oya08}.  We find  approximately  5700
matches with the SDSS catalog.

For  consistency, as  we registered  the optical  images to  the SWIRE
coordinates, thus  we registered the  UKIDSS catalog to our  grid.  We
found a  systematic relative offset  of about $\rm{0.2''}$,  which has
been corrected by  changing the astrometry of the  UKIDSS catalog using
Aladin\footnote{http://aladin.u-strasbg.fr/aladin.gml} \citep{ALADIN}.
We then  matched the optical  and UKIDSS catalogs keeping  the closest
infrared  source  to the  optical  coordinates  within a  $\rm{0.6''}$
radius  to compensate  for  any further  astrometric differences.   We
found approximately  25000 matches between the optical  and the UKIDSS
sources.  From   these  matches,  $98\%$   lie  inside  a   radius  of
$\rm{0.5''}$ from the optical position.
Additionally  $~650$ matches are located
within a  distance between  $\rm{0.5''-0.6''}$ which corresponds  to a
physical distance of 2.5-3 pixels on our images.

\begin{figure}
\includegraphics[width=\linewidth]{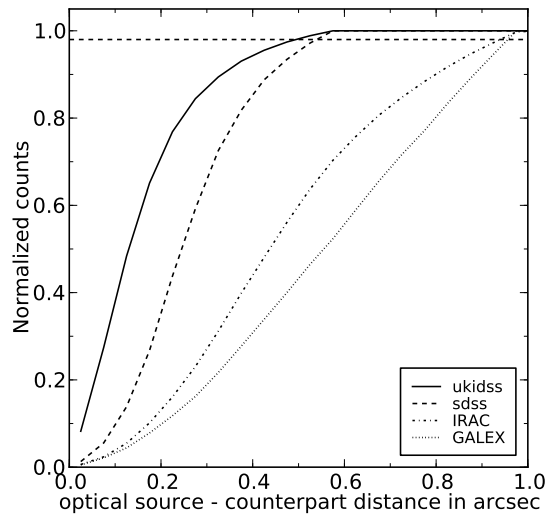}
\caption{Cumulative curve of the distance between the optical position and the counterpart in the UKIDSS (solid line), SDSS (dashed line), IRAC (dot-dashed line), and GALEX (dotted line) catalog.
\label{fig:counterparts}}
\end{figure}

\placefigure{fig:counterparts}

Finally,  we  incorporated   the  Spitzer  catalogs  using  positional
matching within $\rm{1''}$  from the optical position. We  are using a
large  matching  radius  for  the  same  reasons as  in  the  case  of
GALEX. Approximately 26000 sources are matched to an optical source.

In order to  create the final merged catalog, we  keep all the sources
detected in the $\rm{5\sigma}$ level,  and the sources detected in the
$\rm{3\sigma}$ level  which i) have  a UKIDSS counterpart and  ii) are
not  present within  $0.5''$ from  the sources  in  the $\rm{5\sigma}$
catalog.   This  choice  is  a  compromise between  sources  of  lower
significance and false  detections. In the catalog, we  provide a flag
denoting which  catalog each source  originates from. In  addition, in
order to account for problems  originated by blending, we flag sources
that are not isolated. The released catalog consists of 187611 sources
and is described in detail in \S \ref{sec:PhotCat}.

\subsubsection{X-ray Sources and their properties}\label{sec:Xcat}

In the final  catalog, sources that are identified  as counterparts to
the    X-ray   detections   are    flagged.    The    association   in
optical/near-infrared/mid-infrared bands  using the maximum likelihood
ratio technique is described in detail in \citep{Rov11}. We include in
the photometric catalog  389 out of the 409  X-ray detected sources in
the  field. The  20 sources  not present  are either  too faint  to be
detected even  at the $\rm{3\sigma}$ threshold or  are associated with
stars.  For  the  sources  flagged  as  X-ray  detections  we  provide
additional  information that  will be  of crucial  importance  for the
computation  of photometric  redshifts. These  are the  morphology and
variability analysis.

\paragraph{Morphology}
Recently, HST images became available  for the central area of the XMM
region   (covering   in  total   $0.034\;   deg^2$,   red  circle   in
Fig. \ref{fig:area}, PI: Somerville). We used those images to classify
the counterparts of  X-ray sources in i) point-like  and ii) extended,
by visual  inspection. We also  performed morphological classification
by visual  inspection of the B  and Rc bands. The  final morphology is
assigned based primarily on the HST images, complemented by the ground
based images for  the sources with no HST  coverage. We classified 134
sources as point-like and 140  as extended. From the remaining sources
103  are  too  faint to  be  classified  and  12 are  associated  with
saturated sources (see Table \ref{tab:MorphFlags}).

\placetable{tab:MorphFlags}

\paragraph{Variability}
AGN vary  on time scales from  days to years.  This intrinsic property
complicates the  computation of photometric  redshifts. The photometry
is most of the times gathered over many years and a change in flux can
be misinterpreted  as an  emission line. The  variability in  COMBO 17
\citep{Wolf04}  and in  COSMOS \citep{Sal09},  was  quantified through
observations  of the  same energy  band repeated  over the  years. The
comparison of  these repeated observations  allowed the identification
of the variable  sources and the correction of  the photometry. In the
Lockman  Hole, we  can detect  variability but  we cannot  correct for
it. Thus, we limit the  analysis in flagging the variable sources. The
variability  flag will  serve  as warning  for potentially  unreliable
photometric redshifts.
 
We use as proxy of the  variability the flux variation in the $\rm{z'}$ band,
for which we have observations from the Subaru telescope, from LBT and
from  SDSS. The  variability is  expected to  be more  significant for
bright  sources, as  for fainter  sources the  photometric  errors are
large. Taking  into account  only bright sources  ($\rm{18<z<22}$), we
compute  the variability ($Z_{var,k}$,  $k=1,2,3$) and  the associated
error  ($\delta  Z_{var,k}$)  for  all  three pairs of available $\rm{z'}$
filters:
\begin{equation}\label{eq:Dm}
Z_{var,k} = z'_{i} - z'_{j}
\end{equation}
and
\begin{equation}\label{eq:Dmerr}
\delta Z_{var,k} = \sqrt{(z'_{err,i})^2 + (z'_{err,j})^2}
\end{equation}
Using theoretical  magnitudes of galaxies  (see \S \ref{sec:templates}
for  details),  we  verified   that  the  expected  deviation  in  the
photometry, because  of the  slightly differing $\rm{z'}$  filters, is
always less than 0.2 magnitudes for redshifts up to 5.6, for all types
of   galaxies.   Therefore,   we  flag   a  source   as   variable  if
$|Z_{var,k}\pm\delta  Z_{var,k}|>0.2$  in at  least  one  pair of  the
observations.  We  flag 102  sources  as  varying  and 58  sources  as
non-varying. The remaining 229 sources, are either faint ($\rm{z>22}$)
or  they  are  lacking  z-band  photometry and  they  are  treated  as
non-varying.

\subsubsection{Spectroscopic sample}

Compared  to  other  fields,  the  number  of  spectroscopic  redshift
available in  Lockman Hole  is limited and  focused mostly on  AGN and
infrared  galaxies.   From  all  the  available  catalogs  (see  Table
\ref{tab:zref}  for a  complete  list), we  extract the  spectroscopic
redshifts with  the highest confidence. The quality  is provided either
by the authors (and in these cases we cannot verify the quality of the
redshift estimate) or, when the spectrum is available, it is assessed
by us. We  define a  redshift reliable when  more than  1 feature
(either in emission and/or in absorption) is present in the spectrum.

We         are         using         10         redshifts         from
\citet{Schm98} (${0.245<z<2.144}$)   and   50   redshifts   from
\citet{L01} (${0.074<z<4.45}$), where  the sources were observed
as possible counterparts of the X-ray sources detected by ROSAT. Also,
during  the  same observing  run  normal  galaxies  where observed  as
secondary targets. These observations were published in the PhD thesis
of I. Lehmann (Universit\"at Potsdam,  2000). In this way we recovered
additional       29       reliable       spectroscopic       redshifts
(${0.045<z<0.903}$).

Furthermore,   we   include    the   observations   by   \citet{Zap05}
(${0.085<z<1.13}$),   who    studied   the   presence    of   a
superstructure in  the field. Their  catalog contains 48  high quality
spectroscopic  redshifts, with  the superstructure  being  at redshift
$\sim0.8$.   We   include  42   redshifts   from   the  SDSS   catalog
(${0.00<z<3.269}$)  and 44  redshifts retrieved  from NASA/IPAC
Extragalactic Database  \footnote{The NASA/IPAC Extragalactic Database
  (NED)  is  operated by  the  Jet  Propulsion Laboratory,  California
  Institute   of  Technology,   under  contract   with   the  National
  Aeronautics       and       Space       Administration.}       (NED)
(${0.073<z<3.036}$).

Our   group  has   also   observed  sources   in   the  Lockman   Hole
Field.  Focusing  on the  counterparts  of  XMM  detected sources  two
observing runs  were performed with KECK/DEIMOS between  2004 and 2007
(38 high quality redshift, $0.029<z<3.408$) and in 2010
(20  sources,  $0.353<z<1.302$).  Furthermore,  spectroscopic
follow up of  the SWIRE field provided 321  high quality spectroscopic
redshifts (Huang et al. in preparation) ($0.018<z<3.471$).

In total,  we have  602 high quality  spectroscopic redshifts,  out of
which 487  are redshifts  of normal galaxies  with median  redshift of
$z=0.42$  ($0.000<z<3.471$) with  252  galaxies being  inside the  XMM
area.   Furthermore, 115  correspond  to X-ray  detected sources  with
median redshift of $z=0.79$ ($0.024<z<4.45$).

\placetable{tab:zref}

\subsubsection{Description of the Photometric Catalog}\label{sec:PhotCat}

In the  following we give a  brief description of  each column in
  the catalog. We adopt the  value ''-99'' for null fields inside the
catalog. An excerpt of the catalog is presented in Tab. \ref{tab:Photom_cat}.

\begin{itemize}
\item ID -- unique identification number for the final catalog.
\item  Optical coordinates  (J2000)  from the  detection  image --  in
  addition  we provide  the  coordinates of  the  counterparts in  the
  GALEX, UKIDSS, SDSS, and IRAC catalogs.
\item  AB Magnitudes and  errors --  aperture photometry  corrected to
  total (only  for GALEX and  IRAC photometry) for  point-like sources
  and  associated  error  for  the  filters:  $\rm{FUV}$,  $\rm{NUV}$,
  $\rm{U}$, $\rm{B}$, $\rm{V}$, $\rm{z'_{LBT}}$, $\rm{Y}$, $\rm{R_c}$,
  $\rm{I_c}$,  $\rm{z'_{Subaru}}$,  $\rm{u'}$,  $\rm{g'}$,  $\rm{r'}$,
  $\rm{i'}$,   $\rm{z'}$,   $\rm{J}$,   $\rm{K}$,   $\rm{3.6\mu   m}$,
  $\rm{4.5\mu m}$, $\rm{5.8\mu m}$,  $\rm{8\mu m}$.
\item Detection flags -- the meaning of the flags is: 1 - Rc detection
  ($\rm{5\sigma}$), 2 - $\rm{z'}$ detection ($\rm{5\sigma}$), 3 - B detection
  ($\rm{5\sigma}$), 4 - Rc detection ($\rm{3\sigma}$), 3 - $\rm{z'}$ detection
  ($\rm{3\sigma}$), 6 - B detection ($\rm{3\sigma}$).
\item  Photometry  flag  --  complementary  to the  flag  provided  by
  SExtractor indicating  saturation, we masked  problematic regions on
  the images which  include bad pixels and problematic  areas close to
  stars,    using    'Weight    Watcher'    \citep{WW}.    In    Table
  \ref{tab:PhotFlags} we describe the flag values which we provide for
  each  band. For more  details, refer  to the  associated description
  file of the catalog.
\item Neighbor  flag --  sources that have  a neighbor  within $1.5''$
  carry a  flag 1,  while sources without  close by neighbors  carry a
  flag 0.
\item  Star/galaxy classification  -- the  classification  provided by
  SExtractor (1  = star,  0 = galaxy),  measured on  the corresponding
  detection image.
\item Spectroscopic  redshift and  reference -- we  use only  the high
  quality spectroscopic redshifts from the original catalogs according
  to Table \ref{tab:zref}.
\item Variability -- sources with $|Z_{var,k}\pm\delta Z_{var,k}|>0.2$
  carry a flag 1, while the rest carry a flag 0.
\item X-ray detection -- sources detected in the X-rays are flagged as
  1, sources inside  the XMM area without X-ray  detection are flagged
  as 0, while sources outside the XMM area are flagged as -99.
\item X-ray  ID -- the  sources identified as X-ray  counterparts will
  have  the  corresponding  X-ray  ID  number  from  the  XMM  catalog
  \citep{Br08}.  The  non  X-ray  detected  sources have  a  value  of
  XID=-99.
\item Morphology -- as discussed  in detail in \S \ref{sec:Xcat}, this
  column gives  the merged morphology  classification information from
  the HST and ground based  images, provided only for the counterparts
  of the X-ray sources. The flags used and their meaning are presented
  in detail in Table \ref{tab:MorphFlags}.
\end{itemize}

\placetable{tab:Photom_cat}
\placetable{tab:PhotFlags}

\section{Photometric Redshifts}\label{sec:photoz}
In  this section  we  describe the  configuration  of the  photometric
redshift computation. This includes  the SED templates, the extinction
and  redshift  grid,  the  second  order  correction  applied  in  the
photometry as  well as the special  treatment of the  X-ray sample. We
also define the quantities we use later on in the discussion to assess
the quality of the results.

\subsection{LePhare Setup}
We compute  photometric redshifts  for all the  sources in  the field,
using          the          publicly          available          code,
LePhare\footnote{http://www.cfht.hawaii.edu/~arnouts/LEPHARE/lephare.html}. The
code performs least squares  minimization to retrieve the best fitting
template  to the  photometric data.  In order  to achieve  the optimum
result  for both  normal  galaxies  and AGN,  we  treated the  samples
separately and  with different  sets of templates  and priors.  In the
following we  describe the  templates and the  priors used in  the two
cases.

\subsubsection{Templates}\label{sec:templates}

We decided to adopt the  same templates as used by \citet{Il09, Sal09}
in the  COSMOS field.  Being  well tested with a  large spectroscopic
sample, the templates are now  included by default in the distribution
of the LePhare. We briefly summarize here their major characteristics.

For  the normal  galaxies  the template  set  consists of  elliptical (templ. no. 1-7),
spiral (templ. no. 8-19) and  starburst templates (templ. no. 20-31) 
(\citet{Il09},  Fig. 1).  Extinction
according to the Small  Magellanic Cloud (SMC) law \citep{Prevot84} is
applied  for  templates  Sb-SB3,  while  for  the  templates  SB4-SB11
Calzetti  \citep{Cal00} and  modified  Calzetti laws  are applied.  No
additional extinction  is applied for  templates redder than  Sb. The
  intrinsic galactic absorption is computed with values E(B-V) = 0.00,
  0.05,  0.10,  0.15,  0.20,  0.25,  0.30, 0.40,  0.50.  Finally,  the
  templates are calculated at redshifts 0-6 with step $\Delta z=0.01$,
  and at  redshifts 6-7 with  step $\Delta z=0.2$. Emission  lines are
  added  to the  templates  as this  has  been proven  to give  better
  results even in the case of broadband photometry \citep{Il09}.

For the X-ray  detected sample, we are using the  same library used in
\citet{Sal09,  Sal11}  for  computing  the  photometric  redshifts  of
XMM-COSMOS  and  {\it Chandra}-COSMOS.   The  library includes  normal
galaxies, local AGN, and  hybrid templates.  The templates forming the
library  where chosen (and,  in the  case of  the hybrids  created) to
represent  the  spectroscopic  sample  available  for  the  XMM-COSMOS
survey, as documented in  \citet{Sal09}.  The SEDs of galaxies hosting
an AGN  and pure  galaxy differ mostly in the ultra-violet  where the
contribution of the accretion disk  is expected and in the infrared where
the  reprocessed AGN radiation  by the  torus surrounding  the central
black hole is dominant.  The templates of pure AGN dominated sources are taken
from the library  of \citet{Pol07} and are characterized  by a typical
power-law spectrum. To the pure  type 1 AGN, the power-law is extended
to  the UV  beyond  the Ly$\alpha$.  The  reason for  it  is that  the
templates are empirical and thus diminished in the UV by the absorbers
along the  line of  sight. As  LePhare accounts  by default  for this
absorption, without the addition of the power-law, the templates would
be absorbed twice.  More details on the construction  of the templates
can be  found in \citet{Sal09}. We  apply only the  SMC extinction law
with E(B-V) values of 0.00,  0.05, 0.10, 0.15, 0.20, 0.25, 0.30, 0.35,
0.40, 0.45, 0.50.  The templates are calculated with the same redshift
steps as in the case of normal galaxies.

All the sources  of our catalog are also fit with  stellar SEDs of F-K
dwarfs   and   G-K  giant   stars   \citep{Pickles98},  white   dwarfs
\citep{Bohlin95},  low mass  stars  \citep{Chabrier00} and  sub-dwarfs
\cite{Bixler91}.

\subsubsection{Residual zero-point offsets}
\placetable{tab:offsets} 

Systematic  differences can  be found  between templates  and observed
SEDs,  due  to  uncertainties   in  the  templates  and  second  order
calibration  problems in  photometry.  The  impact of  calculating and
using these offsets  on the accuracy of the  result is demonstrated in
\citet{Il06}.  With the option 'AUTO\_ADAPT' in LePhare we compute the
average difference in magnitude between the photometry in a given band
and the photometry for the best SED fitting at the fixed spectroscopic
redshift of  a sample of normal  galaxies ($\rm{18<R_c<24}$, $\sim$260
sources). Iteratively, we then search  for the best set of corrections
that minimize the offsets.  Once found, the offset is applied with the
option  'APPLY\_SYSSHIFT'  to  the  SEDs  when  computing  photometric
redshift for  the entire catalog.   The same offsets are  also applied
when computing  the photometric redshift  for the X-ray  sources.  The
offsets  are presented in  Table \ref{tab:offsets},  and they  are not
included  in  the photometry  presented  in  the  catalog. We  do  not
calculate  any offsets  for the  $\rm{5.8\mu m}$  and  $\rm{8.0\mu m}$
filters of  IRAC, due  to the large  uncertainties in  the theoretical
models in these wavelengths.

We  also include  an additional  factor to  the photometric  errors in
quadrature   (optical  $\rm{0.02\,mag}$,  ultra-violet   and  infrared
$\rm{0.2\,mag}$).  This  factor  compensates  for  the  underestimated
errors provided by SExtractor \citep{Becker07, McCracken01}.

\begin{figure}
\includegraphics[width=\linewidth]{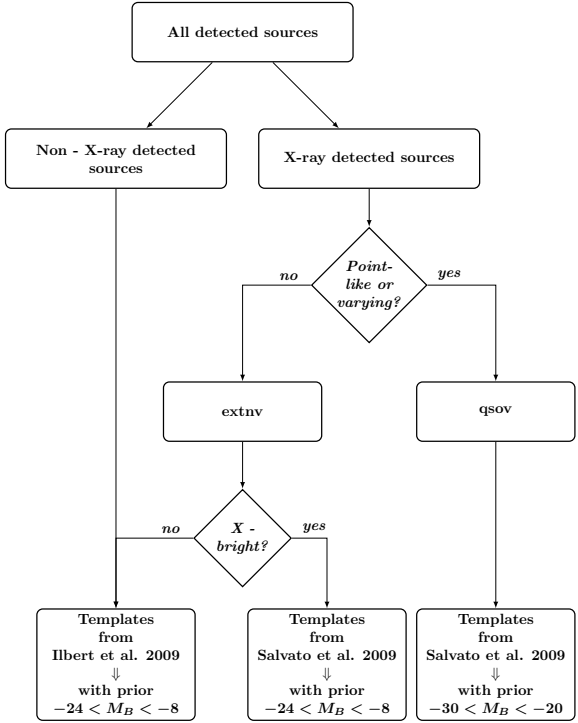}
\caption{The flow chart describes the decision procedure for the optimum combination of templates and priors during the photometric redshift computation. The 'extnv' sample contains the extended and the non-varying sources, while the 'qsov' sample contains the point-like and varying sources. For this work, we adopt $\rm{F_{0.5-2keV}=8\cdot10^{-15}erg\,s^{-1}\,cm^{-2}}$ as flux threshold to separate between X-ray bright and faint sources.
\label{fig:flowchart}}
\end{figure}

\placefigure{fig:flowchart}

\subsubsection{Photometric Redshift Computation}

The separate treatment of the sources based on the X-ray detection and
emission, allows  a pre-selection  of templates and  luminosity priors
that reduces  the possible parametric  space of the  solutions lifting
degeneracies between  the templates and therefore  reducing the number
of wrong  photometric redshift solutions.  In Fig. \ref{fig:flowchart}
we present the flow chart for the proper separation of the sample, and
the template - prior combination we adopt for this work.

First of  all, the  sample is  separated in non - X-ray  detected and
X-ray detected  sources.  For the non  - X-ray detected  sample we use
the  templates  for normal  galaxies  presented  in \citet{Il09}  with
priors  $-24<M_B<-8$, which  is the  typical range  of the  absolute B
magnitude  for normal  galaxies.  For the  X-ray  detected sources,  a
further separation  is required between point-like  or varying sources
(QSOV) and  extended and non-varying (EXTNV) sources.  The QSOV sample
contains the AGN  dominated sources and for them we  are using the AGN
templates of \citet{Sal09} with the same priors $-30<M_B<-20$. 

The EXTNV sample contains  moderately AGN dominated sources, starburst
and normal  galaxies. As demonstrated  in \citet{Sal11}, AGN  that are
low X-ray emitters, extended and  not varying are better fit by normal
galaxy   templates,   even   if   the  X-ray   luminosity   is   above
$\rm{10^{42}erg\cdot   sec^{-1}}$.    For   this  reason,   we   treat
differently  sources   that  above  and   below  a  threshold   set  at
$\rm{F_{0.5-2\,keV}>8\cdot10^{-15}erg\,cm^{-2}\,s^{-1}}$.   Above this
value, the sources are fitted  by AGN templates. The library of normal
galaxies defined by \citet{Il09} is used otherwise. In both cases, the
luminosity prior  $-24<M_B<-8$ is adopted.  It is  important to stress
that  the threshold was  defined using  a sample  of 700  sources with
spectra, belonging to the EXTNV group, in the COSMOS field.

\begin{figure}
\includegraphics[width=\linewidth]{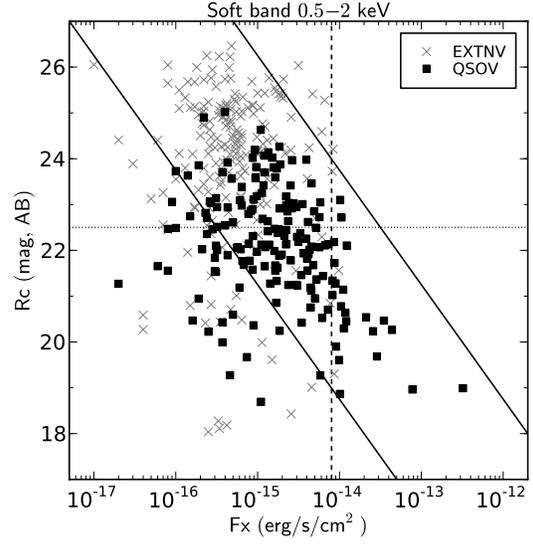}
\caption{Rc magnitude versus  soft X-ray flux $\rm{(0.5-2\,keV)}$. The
  crosses  represent all  the EXTNV  sample and  the squares  the QSOV
  sample. The black  solid lines correspond to $\rm{log(f_X/f_R)\pm1}$
  where the majority  of the AGNs lie. Non-active  galaxies will be in
  the  lower left  region of  the  diagram. The  vertical dashed  line
  denotes  the  X-ray threshold  we  apply  to  the EXTNV  sample  for
  adopting different  templates and priors. The  horizontal line marks
  the bright subsample ($R_c<22.5$). \label{fig:fxfopt} }
\end{figure}

\placefigure{fig:fxfopt}

In Fig.  \ref{fig:fxfopt} we  plot the optical  magnitude ($\rm{R_c}$)
versus  the soft  X-ray  flux ($\rm{0.5-2\,keV}$)  for  all the  X-ray
detected sources.   The squares  indicate sources in the  QSOV sample
and the  crosses indicate sources in  the EXTNV sample.  The two black
solid lines mark  the area in which the  majority of the AGN-dominated
sources is found and they are defined as:
\begin{equation}\label{eq:fxfopt1}
log{(\frac{f_x}{f_{Rc}})}=\pm 1
\end{equation}
where,
\begin{equation}\label{eq:logfopt}
log{(\frac{f_x}{f_{Rc}})}=log{f_x}+\frac{Rc}{2.5}+5.5
\end{equation}
see also \citet{Mac88} \citet{Horn03} and \citet{Brusa05}.

The separation  of the sources  in QSOV and  EXTNV is purely  based on
morphological  and variability  considerations. Examining  sources for
which we  can distinguish  between point-like and  extended morphology
($R_c<24$) we  find that 76.4\% of  the QSOV sources are  found in the
area between the two solid black lines in Fig. \ref{fig:fxfopt}, while
53.5\% of the EXTNV sources are found in the same area, justifying our
original assumption  that QSOV  sources are AGN-dominated  sources and
need to be treated with appropriate templates and luminosity priors.

Focusing on the bright ($Rc<22.5$, dotted line), extended, non varying
sources, we  see that the  above mentioned empirical threshold  in the
X-rays           (vertical           dashed          line           at
$\rm{F_{0.5-2keV}=8\cdot10^{-15}erg\,s^{-1}\,cm^{-2}}$),      separates
efficiently  the  AGN dominated  systems  from  starbursts and  normal
galaxies,  which  populate  the  lower  left corner  of  this  diagram
\citep{Horn03}.  The  positive improvement of the  X-ray threshold when
computing photometric  redshifts for  X-ray sources, is  only marginal
for this work, as the brighter  sources are rare  and the Lockman Hole is  a small
field.  However,  the threshold resulted in a  noticeable improvement on
the XMM-COSMOS surveys \citep{Sal11} and  we decided to adopt the same
strategy for consistency.

\subsection{Outlier-accuracy definition}

\paragraph{Outliers}
Since the  photometric redshift computation is  a multivariate problem
with many  degeneracies, is  it bound to  produce wrong  solutions for
some objects (see for discussion \citet{Rich01}).

Comparing  the  photometric redshift  solutions  to the  spectroscopic
redshifts, we  quantify the fraction  of outliers $\eta$ as  the ratio
$100\cdot  N_{out}/N_{total}$  where  $N_{total}$  is  the  number  of
sources  with  spectroscopic  redshift  and $N_{out}$  the  number  of
sources with:
\begin{equation}\label{eq:eta}
 \frac{|z_{phot}-z_{spec}|}{1+z_{spec}} >0.15
\end{equation}

\paragraph{Accuracy}
The accuracy of the photometric redshifts is usually quantified by the
direct  comparison of  the photometric  redshift solution  against the
spectroscopic  value. In the literature  there are  many  ways  of
computing  the  accuracy, from  the  pure root mean square (rms) of  the  $\Delta z  =
(z_{phot}-z_{spec})/(1+z_{spec})$, to  the rms after 1  or 3 sigma
clipping,  depending on the  authors \citep{Wolf04, Mobasher04, Margoniner08, Dahlen10}.

Another more robust  way to quantify the accuracy of  the sample is to
consider the  median of the  deviations from the  true (spectroscopic)
value. In  this way  the accuracy accounts  also for the  outliers. We
adopted this approach, described in detail in \citet{Hoag83} and used
e.g.     by   \citet{Bram08,  Wuyts08, Il09, Sal09, Car10}.  
The  Normalized Median  Absolute  Deviation (NMAD)  is defined as:
\begin{equation}\label{eq:nmad}
 \sigma_{NMAD}=1.48\cdot median\vert \Delta z\vert
\end{equation}
where $\Delta z  = (z_{phot}-z_{spec})/(1+z_{spec})$. This quantity is
expected to remain close to zero for all redshifts.

\section{Results}\label{sec:results}

\begin{figure}
\includegraphics[width=\linewidth]{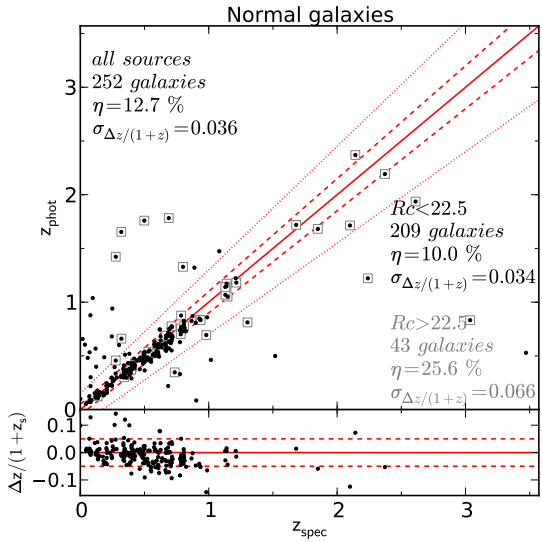}
\caption{Photometric  redshift versus  spectroscopic redshift  for the
  normal  galaxies. With  gray  squares we  denote  the faint  sources
  ($\rm{Rc>22.5\,mag}$).      The      solid      line     is      the
  $\rm{z_{phot}=z_{spec}}$    relation.   The    dashed    lines   are
  $\rm{z_{phot}=0.05\pm   (1+z_{spec})}$.   The   dotted   lines   are
  $\rm{z_{phot}=0.15\pm (1+z_{spec})}$.  Sources that lie  outside the
  dotted lines are defined as outliers.
\label{fig:normalz}}
\end{figure}

\placefigure{fig:normalz}

\subsection{Photometric Redshifts of normal galaxies}

\placetable{tab:photoz}

We  retrieve photometric  redshift  for 184642  non  - X-ray  detected
sources in the Lockman Hole area  with $45\%$ of them computed with at
least  six  photometric  bands.  The  remaining 1978  sources  lack  a
photometric  redshift solution  since they  have only  two photometric
bands available.  In Fig. \ref{fig:normalz} we compare the photometric
versus   the  spectroscopic  redshifts.   For  the   bright  subsample
($\rm{Rc<22.5\,mag}$)    the    accuracy   is    $\sigma_{NMAD}=0.034$
(eq. \ref{eq:nmad}) and the fraction of outliers $\eta=10.0\%$.

\begin{figure}
\includegraphics[width=\linewidth]{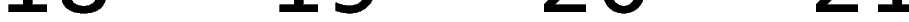}
\caption{The ratio $\rm{(z_{phot}-z_{spec})/(1+z_{spec})}$ versus optical magnitude $\rm{R_c}$ for normal galaxies and corresponding histogram on the right panel. The photometric redshifts of bright sources ($\rm{R_c<22.5}$, black filled circles and black solid line) are mostly confined between $\rm{|(z_{phot}-z_{spec})/(1+z_{spec})|<0.15}$ (dotted lines, marking the outlier region), while the outliers with the largest discrepancies are faint sources ($\rm{R_c>22.5\,mag}$, gray open circles and gray dashed line). \label{fig:Dznormal}}
\end{figure}

Even though our photometric catalog  does not contain medium or narrow
band photometry,  our result does  not suffer from  systematic biases.
This  is  also clear  in  Fig.   \ref{fig:Dznormal},  where the  ratio
$(z_{phot}-z_{spec})/(1+z_{spec})$ is plotted as a function of optical
magnitude.        The       distribution       of      the       ratio
$(z_{phot}-z_{spec})/(1+z_{spec})$  is  centered  at zero  (histogram,
right panel) both for bright  ($R_c<22.5$, black solid line) and faint
sources ($R_c>22.5$, gray dashed  line).  For the latter, however, the
fraction   of  outliers  is higher  and  consequently   the  accuracy
decreases.  This is  a well  known trend,  discussed already  by many
other authors  ( \cite[e.g.,][]{Car10, Barro11, Il09, Sal09}; 
at  fainter  magnitudes  the spectral  energy
distribution is less tightly constrained,  and only by upper limits in
some bands, or has large statistical uncertainties associated with the
photometry.

\placefigure{fig:Dznormal}

\begin{figure}
\includegraphics[width=\linewidth]{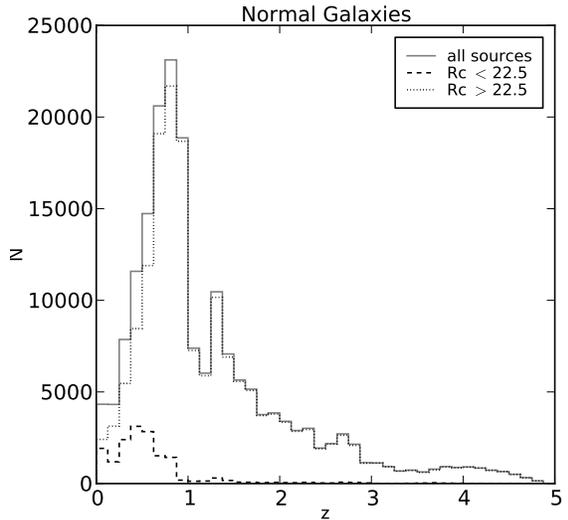}
\caption{Redshift distribution for normal galaxies, using spectroscopic redshift when available. Bright galaxies ($\rm{R_c<22.5\,mag}$) lie mostly at redshift $z<1$. \label{fig:histonormal} }
\end{figure}

In Table \ref{tab:photoz} we summarize the accuracy and the percentage
of the  outliers for various  subsamples. The majority of  the sources
with spectra  available have redshift $z_{\rm  spec}<$1, and therefore
we  can characterize  the  quality  of our  redshift  solutions, in  a
statistically robust  way, only for low redshift  sources. For $z_{\rm
  spec}<$1, the accuracy  is $\sigma_{NMAD}=0.034$ (eq. \ref{eq:nmad})
and    the    fraction   of    outliers    is   $\eta=9.6\%$.     Fig.
\ref{fig:histonormal}  shows the redshift  distribution of  the normal
galaxies    (solid    line)     separated    in    optically    bright
($\rm{Rc<22.5\,mag}$,    dashed     line)    and    optically    faint
($\rm{Rc>22.5\,mag}$,  dotted  line),  where  spectroscopic  redshifts
substitute photometric redshifts when possible.

\placefigure{fig:histonormal}

\begin{figure}
\includegraphics[width=\linewidth]{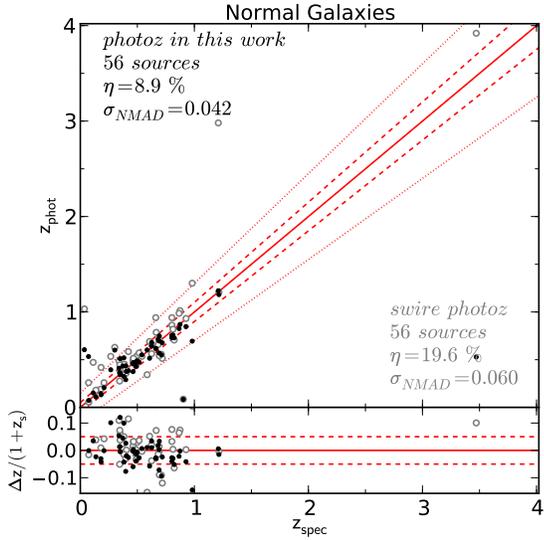}
\caption{Comparing the quality of new photometric redshifts, using sources in common in the present work (black filled dots) and the SWIRE catalog (gray open circles). There is a dramatic decrease in the fraction of outliers and the accuracy is improved.
\label{fig:swirenormal} }
\end{figure}

\placefigure{fig:swirenormal}

In   Fig.  \ref{fig:swirenormal}  we  also  compare  the  photometric
redshifts  from this  work with  the previous  available photometric
redshifts available from \citep{Row08}.  We consider only the sources  in common to
the two samples for which spectroscopic redshifts are available (56 sources). Due
to the  increased number of bands  used in this work,  we achieve an
accuracy improved up to a factor of 1.5 for normal galaxies and 2 times less outliers.

\subsection{Photometric redshifts of X-ray detected sources}

\begin{figure}
\includegraphics[width=\linewidth]{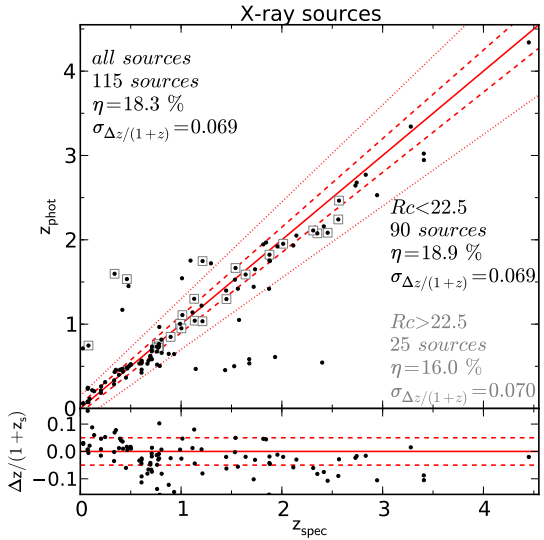}
\caption{Photometric redshift versus spectroscopic redshift for the X-ray detected sources. The symbols and lines are the same as in Fig. \ref{fig:normalz}.
\label{fig:Xrayz}}
\end{figure}

\begin{figure}
\includegraphics[width=\linewidth]{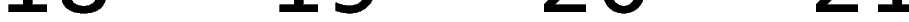}
\caption{The ratio $\rm{(z_{phot}-z_{spec})/(1+z_{spec})}$ versus optical magnitude $\rm{R_c}$ for X-ray sources and corresponding histogram on the right panel. The squares denote the QSOV sample (point-like or varying sources) and the crosses denote the EXTNV sample (extended and not varying sources). The variability is a key factor in the photometric redshift computation, as 11 out of the 16 outliers belonging in the QSOV sample are identified as varying sources. As in the case of the normal galaxies, the largest discrepancies are found for faint sources ($\rm{R_c>22.5\,mag}$). 
\label{fig:Dzactive} }
\end{figure}

\placefigure{fig:Xrayz}
\placefigure{fig:Dzactive}

\begin{figure}
\includegraphics[width=\linewidth]{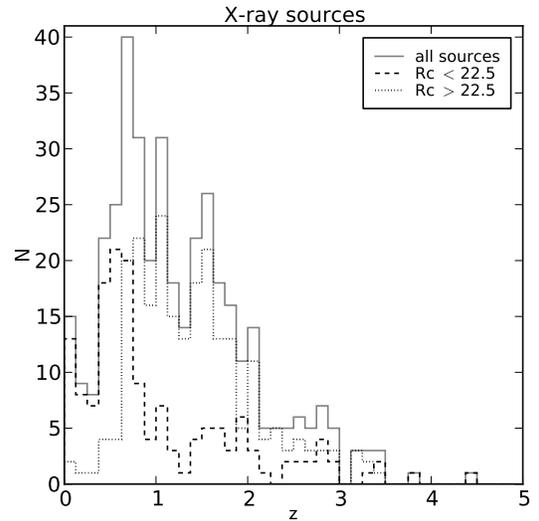}
\caption{Redshift distribution of X-ray sources, using spectroscopic redshift when available. \label{fig:histoX} }
\end{figure}

\placefigure{fig:histoX}

For the counterparts of the X-ray sources, mostly AGN, the accuracy of
the bright  subsample is $\sigma_{NMAD}=0.069$, while  the fraction of
outliers is $\eta=18.9\%$ (Fig.  \ref{fig:Xrayz}).  This is similar to
the accuracy  reached by XMM-COSMOS and {\it  Chandra}-COSMOS when the
same  bands and  depths  available for  Lockman  Hole are  considered,
without correcting for variability.   The impact of variability in our
photometric redshift computation is shown in Fig.  \ref{fig:Dzactive}.
Even    thought   the    distribution    of   $(z_{\rm    phot}-z_{\rm
  spec}/(1+z_{spec})$ is peaked around zero, there are some sources in
the  QSOV  subsample  (squares)  that  show  quite  large  deviations.
Indeed,  11 out  of the  16 outliers  are flagged  as  varying sources
indicating  once  more the  care  that  should  be taken  in  planning
photometric  observations  of  AGN.   In  Table  \ref{tab:Xphotoz}  we
summarize our results separating them  on the basis of the classification
of  the  sources   (either  EXTNV  or  QSOV)  and   brightness  .   In
Fig. \ref{fig:histoX}  we present the  histogram of the  redshifts for
the X-ray detected sources  (solid line) separated in optically bright
($\rm{Rc<22.5\,mag}$,    dashed     line)    and    optically    faint
($\rm{Rc>22.5\,mag}$, dotted line).

In  Table \ref{tab:photoz}  we  give the  detailed  evaluation of  our
results, in direct  comparison with the non X-ray  detected sample. As
it is expected, the photometric redshifts of the brightest sources are
more accurate than the redshifts  for the fainter sources. The highest
accuracy is  reached at $z<1$ with  $\sigma_{NMAD}=0.066$ and fraction
of outliers $\eta=16.4\%$.

\begin{figure}
\includegraphics[width=\linewidth]{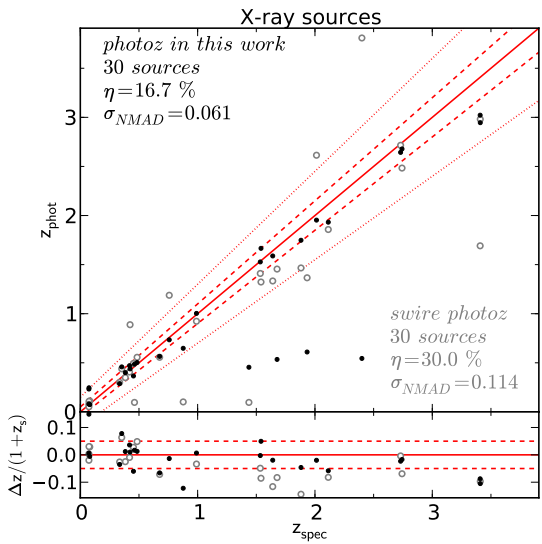}
\caption{Comparison between previous photometric redshifts and results from this work for X-ray detected sources. The symbols are the same as in Fig. \ref{fig:swirenormal}. As in the case of normal galaxies, the outliers are reduced significantly and the accuracy is improved.
\label{fig:swireX} }
\end{figure}

\placetable{tab:Xphotoz}
\placefigure{fig:swireX}

In  Fig. \ref{fig:swireX} we  compare our  results to  the photometric
redshifts of \citet{Row08}. Apart from the increased number of used bands
an  additional  reason  for  the  improvement  is  that,  contrary  to
\citet{Row08}, our work was tuned to this kind of sources. Rather than
adding AGN-dominated  templates to the library of  normal galaxies, we
limited  the degeneracies  by using  only AGN-dominated  templates and
appropriate priors, thus achieving more accurate results by a factor of 
1.8 both in terms of accuracy and outliers.\\

\begin{figure}
\includegraphics[width=\linewidth]{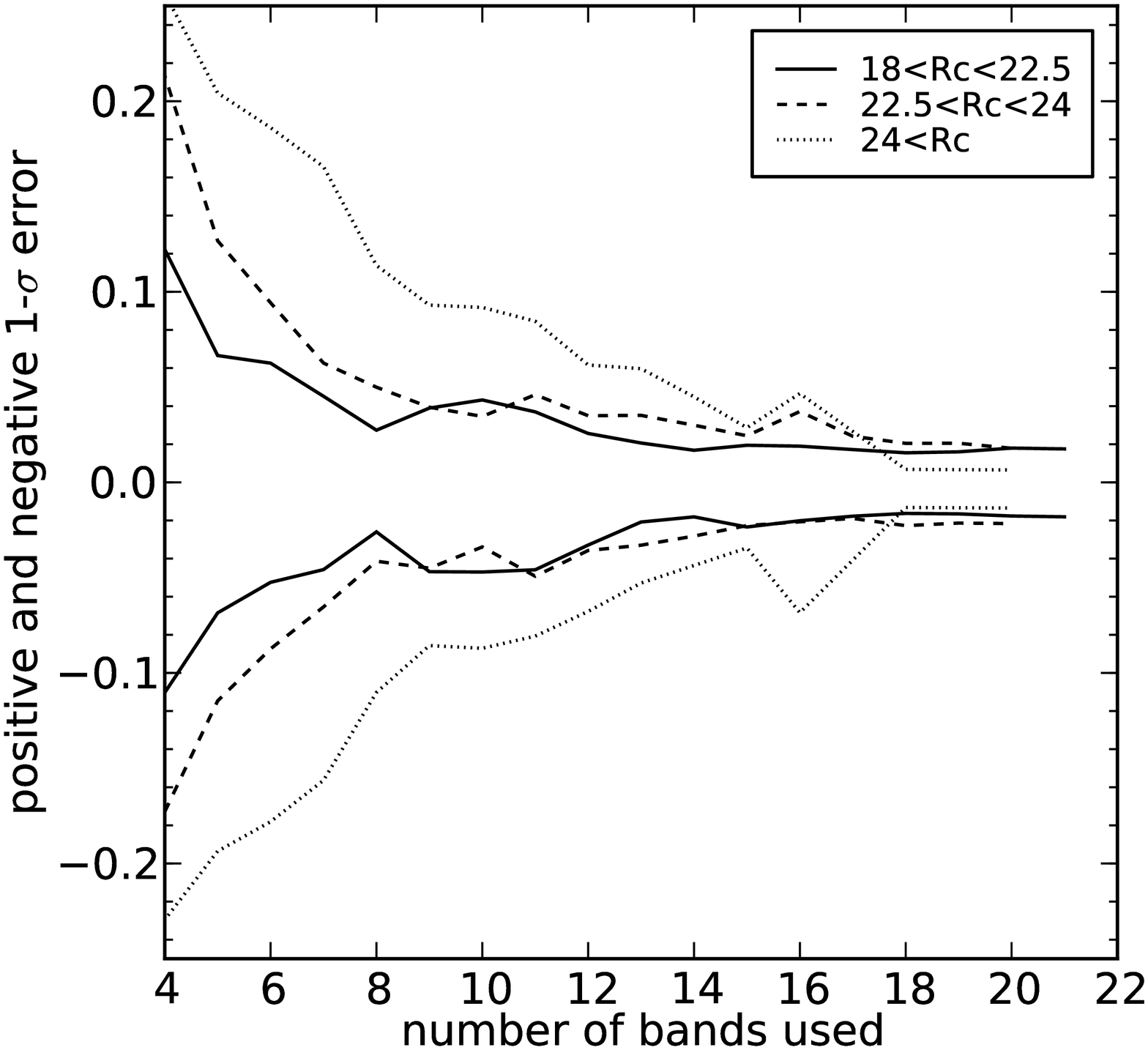}
\caption{Mean error of photometric redshifts versus the number of bands used during the fitting. The bright sample (solid line) has a consistent behavior when six or more number of bands are used. The faint sample (dashed line) approaches the accuracy of the bright sample in cases when eight bands or more are available. \label{fig:errornormal}}
\end{figure}

\placefigure{fig:errornormal}

As already discussed, the  accuracy of photometric redshifts correlates
with the faintness of the  sources.  As a consequence, when the source
is    faint,     less    photometry    is     also    available.
In Fig. \ref{fig:errornormal} we plot the median positive and negative
$1-\sigma$ errors, defined as  ($z_{best68,high} - z_{best}$) and ($z_{best}
- z_{best68,low}$), per number of bands  used in the SED fitting for all sources
(non  X-ray and  X-ray  detected sources).   As  expected, the  bright
sources show small errors even when  using only a few bands, as bright
sources have  usually small  errors associated with  their photometry,
thus making the least squares fitting more precise.  We underline that
narrow errors of photometric redshifts does not mean that the solution
is the  correct one,  rather that the  probability distribution  has a
narrow peak around a given value.

\subsection{Star/galaxy separation}\label{sec:star}

\begin{figure}
\includegraphics[width=\linewidth]{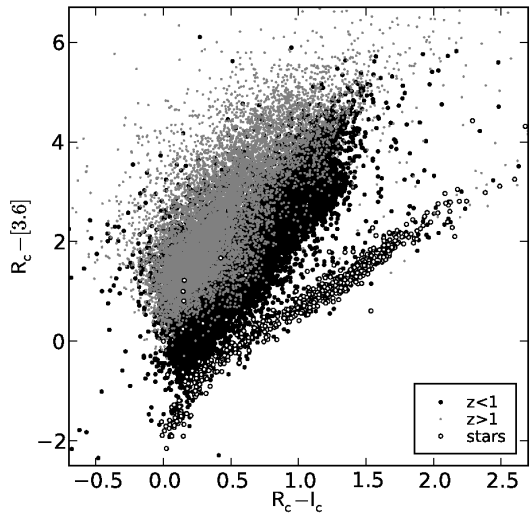}
\caption{Color - color plot demonstrating the star - galaxy separation. Gray dots indicate high redshift galaxies, black circles low redshift galaxies and open circles indicate the stars. Most of the identified stars through the SED fitting, lie in the expected locus for stars. \label{fig:stars} }
\end{figure}

\placefigure{fig:stars}

Rather  than  distinguishing between  stars  and galaxies,  SExtractor
separates the objects in point-like and extended. This is performed by
comparison of the measured FWHM of an object with the PSF of the image
(given  as input),  as  a result  the  code is  unable to  distinguish
between stars and unresolved galaxies.

 To  assess the  limitations of  SExtractor we  compare the  number of
 stars defined by the code,  with the expected number of stars defined
 by                         stellar                         population
 synthesis \footnote{http://model.obs-besancon.fr/} models of galactic
 stars in  the area covered  by Lockman Hole.   In the case  of bright
 objects ($\rm{R_c}<19$)  the number  of stars detected  by SExtractor
 (90  sources) agrees  well with  the expected  number  (103 sources).
 Inversely,  in the  case  of less  bright objects  ($20<\rm{R_c}<22$)
 SExtractor  classifies $\sim  1500$  sources as  stars, almost  three
 times more the number predicted by the simulation ($\sim 440$ stars),
 as it confuses point-like  sources with unresolved ones.  
 
At  the  same time,  SED  fitting  can  misclassify objects  as  stars
especially  in  the  case  of  elliptical galaxies  that  are  lacking
infrared photometry. To compensate between the two effects, we flag as
stars sources  that have been  identified as stars both  by SExtractor
($\rm{Classification}>0.95$)     and     by     the    SED     fitting
($\rm{2\cdot\chi^2_{star}<\chi^2_{best}}$)   using  star   templates  from
\citet{Pickles98,Bohlin95,Chabrier00,Bixler91}.        With       this
conservative approach we flag $\sim  700$ sources as stars , excluding
most  of   the  false  identifications.   Indeed,   according  to  the
simulation we expect $\sim 800$  stars in the Lockman Hole area having
the same magnitude distribution as the the sources identified as stars
from the previous criterion.

Fig. \ref{fig:stars} shows a color-color plot of stars (open circles),
low  redshift (black)  and  high redshift  (gray)  galaxies. With  the
combined criterion  of morphology and  SED fitting, we  retrieve stars
mainly in  the expected locus according to  theoretical templates (see
also \citet{Il09}).

\subsection{Description of the Photometric Redshift Catalog}\label{sec:zcat}
In the  following we  give a description  of the  photometric redshift
catalog, which we provide  separately from the photometry catalog. An
excerpt  of the  catalog is  given in  Tab.  \ref{tab:Photoz_cat}. The
photometric redshift catalog includes:
\begin{itemize}
\item ID -- corresponding identification number from the photometric catalog.
\item $z_{best}$ -- the best fitted solution for the photometric redshift.
\item $z_{best68,low}$ -- the lowest redshift at 68\% significance.
\item $z_{best68,high}$ -- the highest redshift at 68\% significance.
\item $z_{best90,low}$ -- the lowest redshift at 90\% significance.
\item $z_{best90,high}$ -- the highest redshift at 90\% significance.
\item $\chi^2_{best}$ -- the lowest $\chi^2$ value for the best fitted galaxy model.
\item $PDZ_{best}$ -- the probability that $z_{best}$ is the correct photometric redshift.
\item  $model_{best}$ -- the  number corresponding  to the  model best
  fitting  the SED.  100+(1,  \ldots,  31) from \citet{Il09}; 1,
  \ldots, 30 from \citet{Sal09}.
\item $Ext-law_{best}$ -- the extinction-law applied for computing $z_{best}$.
\item $E(B-V)_{best}$ -- the absorption  applied for computing $z_{best}$.
\item $Nband_{best}$ -- number of bands used in the SED fitting.
\item[] Similarly for the second photometric redshift solution, when available:
\item $z_{sec}$ -- the second best solution for the photometric redshift.
\item $\chi^2_{sec}$ -- the corresponding  $\chi^2$ value .
\item $PDZ_{sec}$ -- the probability that the second $z_{best}$ is the correct photometric redshift.
\item $model_{sec}$ -- the  number corresponding  to the  model best
  fitting  the SED.
\item $E(B-V)_{sec}$ -- the absorption  applied for computing the second $z_{best}$:
\item $\chi^2_{star}$ -- the lowest $\chi^2$ value for the best fitted star model.
\item $Flag_{star}$ -- we flag stars (1 = star, 0 = galaxy), as discussed in \ref{sec:star}
\end{itemize}

\placetable{tab:Photoz_cat}

\section{Conclusions}\label{sec:conclusions}

We  present and  publicly release  an optically  based multiwavelength
photometric catalog  which contains 21  bands for sources  detected in
the Lockman  Hole area.   The catalog contains  187611 sources  out of
which  389 sources  are  associated with  X-ray  sources.  The  $50\%$
detection   limits    ($\rm{5\sigma}$)   for   the    photometry   are
$\rm{R_c=26.1\,mag}$, $\rm{z'=24.8\,mag}$  and $\rm{B=27\,mag}$.  This
is the first public catalog containing deep multiwavelength photometry
for the Lockman Hole Deep Field.

Even  though the  lack of  narrow/medium band  photometry  will affect
studies that  require detailed information  of the SED of  the source,
the collection of broad-band  photometric measurements we present here,
can serve as a first educated guess  of the shape of the SED, which we
will address, focusing on the AGN in future work.

We  also present  and publicly  release a  complementary  catalog with
photometric  redshift information  for all  sources. Depending  on the
nature of the sources (non X-ray and X-ray detected) we used different
templates  and  priors,  allowing  a  final accuracy  for  the  bright
subsample  ($\rm{R_c<22.5\,mag}$)   of  $\sigma_{NMAD}=0.034$  with  a
fraction   of   $\rm{10\%}$   outliers   for   normal   galaxies   and
$\sigma_{NMAD}=0.069$  with a fraction  of $\rm{18.9\%}$  outliers for
X-ray detected sources.

\acknowledgments  We  gratefully   acknowledge  the  contributions  of
G. Szokoly,  Alberto Franceschini,  Lucia Marchetti. We  thank Stefano
Berta for  reading the manuscript  and providing helpful  comments. We
acknowledge the anonymous  referee for providing constructive comments
and suggestions.  MS and GH acknowledge support by the German Deutsche
Forschungsgemeinschaft, DFG Leibniz Prize (FKZ HA 1850/28-1).

{\it Facilities:} \facility{KECK}, \facility{HST}, \facility{VLT}, \facility{XMM}, \facility{LBT}, \facility{Subaru}, \facility{SDSS}.



\clearpage

\clearpage
\begin{deluxetable}{lccccccccccc}
\tablecolumns{9}
\tablewidth{0pc}
\tabletypesize{\scriptsize}
\setlength{\tabcolsep}{0.03in} 
\tablecaption{Properties of recent X-ray fields sorted in decreasing area.\label{tab:fields}}
\tablehead{
\colhead{} & \colhead{}& \colhead{Average}  & \colhead{Number} & \multicolumn{4}{c}{Flux limit} & \colhead{X-ray}\\
\colhead{Field} & \colhead{Area} & \colhead{Exposure} & \colhead{of} & \multicolumn{4}{c}{} & \colhead{point source}\\
\cline{5-8}
\colhead{} & \colhead{($\rm{deg^2}$)} & \colhead{(ks)} & \colhead{sources} & \colhead{$0.5-2\rm{keV}$} & \colhead{$2-8\rm{keV}$} &\colhead{$2-10\rm{keV}$} & \colhead{$5-10\rm{keV}$} & \colhead{catalog} }
\startdata
XMM-COSMOS & 2.13 & 60 & 1887 & $1.7\cdot10^{-15}$ &  -   & $9.3\cdot10^{-15}$ & $1.3\cdot10^{-14}$ & \citet{Cappelluti07} \\ 
Chandra-COSMOS  & 0.9 & 200 & 1761 & $1.9\cdot10^{-16}$ &  -   &  $7.3\cdot10^{-16}$ &  & \citet{Elvis09} \\ 
AEGIS & 0.67 & 200 & 1325 & $5.3\cdot10^{-17}$ & -& $3.8\cdot10^{-16}$ & -& \citet{Laird09} \\
ECDFS & 0.3 & 250 &  762 & $1.1\cdot10^{-16}$ & $6.7\cdot10^{-16}$ & -& -& \citet{Lehmer05} \\
{\bf Lockman Hole} & {\bf 0.2} & {\bf 185} &{\bf 409} & $\mathbf{1.9\cdot10^{-16}}$ & {\bf-} & $\mathbf{9\cdot10^{-16}}$ & $\mathbf{1.8\cdot10^{-15}}$ & {\bf \citet{Br08}}\\
CDFN & 0.124 & 1000\tablenotemark{a} & 503 & $2.7\cdot10^{-17}$ & $1.4\cdot10^{-16}$ & -& -& \citet{Alex03}\\
CDFS & 0.129 & 2000\tablenotemark{a} & 740 & $9.1\cdot10^{-18}$ & $5.5\cdot10^{-17}$ & -&- & \citet{Xue11} \\
\enddata
\tablenotetext{a}{$50\%$ of the field has higher exposure than the quoted number.}
\end{deluxetable}

\begin{deluxetable}{lrrrrrrrrr}
\tablecolumns{10}
\tabletypesize{\scriptsize}
\tablewidth{0pc}
\tablecaption{Observational facts for the ground-based observations \label{tab:LimMag}}
\tablehead{
\colhead{} & \multicolumn{5}{c}{Large Binocular Telescope} & \colhead{} & \multicolumn{3}{c}{Subaru}\\
\cline{2-6} \cline{8-10}
\colhead{} & \colhead{U} & \colhead{B} & \colhead{V} &\colhead{z'} & \colhead{Y} & \colhead{} & \colhead{Rc} & \colhead{Ic} & \colhead{z'}}
\startdata
Area ($\rm{deg^2}$) & 0.26 & 0.25 & 0.19 & 0.20 & 0.19 & & 0.53 & 0.53 & 0.53 \\
Filter $\rm{\lambda_{eff}(\AA)}$ & 3573 & 4249 & 5405 & 9050 & 9880 & & 6518 & 7957 & 9064 \\
Filter FWHM ($\rm{\AA}$) & 540 & 916 & 845 & 1053 & 465 & & 1167 & 1381 & 1154 \\
Exposure (sec) & 49680 & 19972 & 9540 & 14400 & 10980 & & 3920 & 6235 & 10400\\
PSF FWHM ($\rm{''}$) & 1.06 & 0.90 & 0.95 & 1.06 & 0.60 & & 0.90 & 0.98 & 0.96\\
Limiting Magnitude (5$\sigma$, AB) & 26.7 & 27.0 & 26.7 & 24.2 & 23.5 & & 26.1 & 25.5 & 24.8\\
AB correction\tablenotemark{a} & 0.964 & -0.046 & -0.005 & 0.528 & 0.558 & & 0.207 & 0.436 & 0.521\\
\enddata
\tablenotetext{a}{As computed by LePhare: $M_{AB}$=$M_{Vega}$+$AB_{corr}$}
\end{deluxetable}

\begin{deluxetable}{ccccc}
\tablecolumns{4}
\tablewidth{0pc}
\tabletypesize{\scriptsize}
\tablecaption{Morphology's flag description \label{tab:MorphFlags}}
\tablehead{\colhead{} & \colhead{} & \multicolumn{2}{c}{Number of Sources}\\
\cline{3-4}
\colhead{Description} & \colhead{Flag} & \colhead{HST} & \colhead{Ground based} & \colhead{Final morphology}}
\startdata
too faint / unresolved source & -2\phn & 44 & 86 & 103\\
photometry blended with nearby source & -1\phn & 5 & 10 & 12\\
extended source & \phs0\phn & 20 & 151 & 140 \\ 
point-like source & \phs1\phn & 23 & 123 & 134\\ 
\enddata
\end{deluxetable}

\begin{deluxetable}{rl}
\tablecolumns{2}
\tabletypesize{\scriptsize}
\tablewidth{0pc}
\tablecaption{Spectroscopic redshift reference \label{tab:zref}}
\tablehead{\colhead{Flag} & \colhead{reference}}
\startdata
1 & \citet{L01}\\
2 & Lehmann, PhD Thesis, Potsdam (2000) \\
3 & unpublished redshift KECK/DEIMOS (2004-2007)\\
4 & unpublished redshift KECK/DEIMOS (2010)\\
5 & SDSS DR2\\
6 & \citet{Zap05} \\
7 & \citet{Chap05}\\
8 & \citet{Schm98}\\
9 & \citet{Swin04}\\
10 & \citet{Chap04}\\
11 & \citet{Hain09}\\
12 & \citet{Bar04}\\
13 & \citet{Stevens03}\\
14 & \citet{Mainieri02}\\
15 & \citet{Rod05}\\
16 & \citet{Oyabu05}\\
17 & \citet{Mateos05}\\
18 & \citet{Hasinger98}\\
19 & \citet{Ish01}\\
20 & \citet{Smail04}\\
21 & \citet{Ciliegi03}\\
22 & \citet{Hashim05}\\
23 & \citet{Ivison05} \\
24 & \citet{Fadda02} \\
25 & Huang et al. (in preparation)\\
26 & \citet{Afonso11})\\
\enddata
\end{deluxetable}

\begin{deluxetable}{cccccccccccccc}
\tablecolumns{14}
\tablewidth{0pc}
\tabletypesize{\tiny}
\setlength{\tabcolsep}{0.04in} 
\tablecaption{Lockman Hole Photometric catalog \label{tab:Photom_cat}}
\tablehead{\colhead{ID} & \colhead{ra} & \colhead{dec} &\colhead{mag} & \colhead{mag} & \colhead{mag} & \colhead{neigh} & \colhead{classification} & \colhead{zspec} & \colhead{zspec} & \colhead{var} & \colhead{XMM} & \colhead{XMMNR} & \colhead{morphology}\\
\colhead{ } & \colhead{ } & \colhead{ } &\colhead{ } & \colhead{err} & \colhead{flag} & \colhead{flag} & \colhead{ } & \colhead{ } & \colhead{cat} & \colhead{ } & \colhead{detection} & \colhead{ } & \colhead{ }}
\startdata
  134555 & 162.5407563 & 57.64900553 & 25.6317 & 0.057156 & 1 & 0 & 0.40040278 & -99.0 & -99 & -99 & -99 & -99 & -99\\
  144534 & 162.5407569 & 57.70602821 & 24.6622 & 0.03174 & 0 & 0 & 0.5890769 & -99.0 & -99 & -99 & -99 & -99 & -99\\
  43512 & 162.5407670 & 57.12902011 & 25.3024 & 0.062409 & 0 & 0 & 0.37800178 & -99.0 & -99 & -99 & -99 & -99 & -99\\
  40455 & 162.5407798 & 57.11082514 & 20.6579 & 0.00478 & 0 & 0 & 0.4363798 & -99.0 & -99 & -99 & -99 & -99 & -99\\
  41535 & 162.5407868 & 57.11838781 & 25.1448 & 0.052445 & 0 & 0 & 0.5059889 & -99.0 & -99 & -99 & -99 & -99 & -99\\
\enddata
\tablenotetext{Note}{Only a portion of this table is shown here for guidance regarding its form and content. Table \ref{tab:Photom_cat} is published in its entirety in the electronic edition of \apjs . For a detailed description see \S \ref{sec:PhotCat}.}
\end{deluxetable}

\begin{deluxetable}{rl}
\tablecolumns{2}
\tabletypesize{\scriptsize}
\tablewidth{0pc}
\tablecaption{Photometry's flag description \label{tab:PhotFlags}}
\tablehead{\colhead{Flag} & \colhead{Description}}
\startdata
-99 & 99 value produced by SExtractor either for \\
\phn & the magnitude or the error,\\
\phn &  or source outside of the field \\
-5\phn & flag in the UKIDSS catalog\\
\phn & marking noise in the JHK bands \\
-4\phn & saturation or incomplete/corrupted data \\
\phn & produced by SExtractor \\
-3\phn & source inside a stripe, as marked by optical\\
\phn & inspection of the images \\
-2\phn & magnitude error was negative \\
-1\phn & magnitude error was greater than 1 \\
\phs0\phn & everything is OK \\
\phs1\phn & FWHM in the detection band is zero \\
\phs2\phn & source inside wings of stars, potentially fake \\
\phs3\phn & magnitude greater than the detection limit \\ 
\enddata
\end{deluxetable}

\begin{deluxetable}{cc}
\tablecolumns{2}
\tabletypesize{\scriptsize}
\tablewidth{0pc}
\tablecaption{Offsets between theoretical templates and observations \label{tab:offsets}}
\tablehead{\colhead{Filter} & \colhead{Offset\tablenotemark{a}} }
\startdata
FUV (GALEX) & -0.165\\
NUV (GALEX) & -0.356 \\
U (LBT) & 0.125 \\
B (LBT) & -0.026\\
V (LBT) & -0.038\\
z' (LBT) & -0.054\\
Y (LBT) & 0.141\\
$\rm{R_c}$ (Subaru) & -0.037 \\
$\rm{I_c}$ (Subaru) & -0.020\\
z' (Subaru) & -0.004\\
u' (SDSS) & 0.271\\
g' (SDSS) & -0.115\\
r' (SDSS) & 0.006\\
i' (SDSS) & 0.104\\
z' (SDSS) & 0.083\\
J (UKIRT) & 0.249\\
K (UKIRT) & 0.294\\
$\rm{3.6\mu m}$ (IRAC) & 0.203\\
$\rm{4.5\mu m}$ (IRAC) & 0.346\\
\enddata
\tablenotetext{a}{Included in the computation of the photometric redshifts, but not in the released catalog.}
\end{deluxetable}

\begin{deluxetable}{lccccccc}
\tablecolumns{8}
\tabletypesize{\scriptsize}
\tablewidth{0pc}
\tablecaption{Photometric redshift accuracy \label{tab:photoz}}
\tablehead{
\colhead{} & \multicolumn{3}{c}{Non X-ray sources} & \colhead{} & \multicolumn{3}{c}{X-ray sources}\\
\cline{2-4} \cline{6-8}
\colhead{} &  \colhead{N(zspec)} &\colhead{$\sigma$} & \colhead{$\eta (\%)$} & \colhead{} & \colhead{N(zspec)} &\colhead{$\sigma$} & \colhead{$\eta (\%)$}}
\startdata
all sources & 252 & 0.036 & 12.7 && 115 & 0.069 & 18.3 \\
\hline
$R_c<22.5$ & 209 & 0.034 & 10.0 && 90 & 0.069 & 18.9 \\
$R_c>22.5$ & 43 & 0.066 & 25.6 && 25 & 0.070 & 16.0 \\ 
\hline
$0<z<1$ & 230 & 0.034 & 9.6 && 67 & 0.066 & 16.4 \\
$1.0<z<5$ & 22 & 0.146 & 45.5 && 48 & 0.078 & 20.8 \\
\enddata
\end{deluxetable}

\begin{deluxetable}{lccccccc}
\tablecolumns{8}
\tablewidth{0pc}
\tabletypesize{\scriptsize}
\tablecaption{Detailed photometric redshift accuracy for the X-ray detected sample\label{tab:Xphotoz}}
\tablehead{
\colhead{} & \multicolumn{3}{c}{QSOV} & \colhead{} & \multicolumn{3}{c}{EXTNV}\\
\cline{2-4} \cline{6-8}
\colhead{} &  \colhead{N(zspec)} &\colhead{$\sigma$} & \colhead{$\eta (\%)$} & \colhead{} & \colhead{N(zspec)} &\colhead{$\sigma$} & \colhead{$\eta (\%)$}}
\startdata
all sources & 85 & 0.071 & 18.8 && 30 & 0.056 & 16.7 \\
\hline
$R_c<22.5$ & 70 & 0.084 & 21.4 && 20 & 0.036 & 10 \\
$R_c>22.5$ & 15 & 0.042 & 6.7 && 10 & 0.126 & 30.0 \\
\enddata
\end{deluxetable}

\begin{deluxetable}{ccccccccccccccccccc}
\tablecolumns{19}
\tablewidth{0pc}
\setlength{\tabcolsep}{0.025in} 
\tabletypesize{\tiny}
\tablecaption{Lockman Hole Photometric Redshift catalog \label{tab:Photoz_cat}}
\tablehead{\colhead{IDENT} & \colhead{Z\_BEST} & \colhead{Z\_BEST68} &\colhead{Z\_BEST68} & \colhead{Z\_BEST90} & \colhead{Z\_BEST90} & \colhead{CHI} & \colhead{PDZ} & \colhead{MOD} & \colhead{EXTLAW} & \colhead{EBV} & \colhead{NBAND} & \colhead{Z} & \colhead{CHI} & \colhead{PDZ} & \colhead{MOD} & \colhead{EBV} & \colhead{CHI} & \colhead{STAR}\\
\colhead{ } & \colhead{} & \colhead{LOW} &\colhead{HIGH} & \colhead{LOW} & \colhead{HIGH} & \colhead{BEST} & \colhead{BEST} & \colhead{BEST} & \colhead{BEST} & \colhead{BEST} & \colhead{USED} & \colhead{SEC} & \colhead{SEC} & \colhead{SEC} & \colhead{SEC} & \colhead{SEC} & \colhead{STAR} & \colhead{FLAG}}
\startdata
  134555 & 1.6691 & 0.57 & 1.98 & 0.48 & 2.36 & 2.16843 & 42.222 & 31 & 0 & 0.0 & 3 & 0.63 & 2.80791 & 16.042 & 31 & 0.0 & 0.660721 & 0\\
  144534 & 0.7756 & 0.76 & 0.79 & 0.73 & 0.8 & 92.5442 & 88.25 & 13 & 1 & 0.4 & 3 & 1.49 & 97.5873 & 6.911 & 4 & 0.0 & 53.7372 & 0\\
  43512 & 1.6959 & 1.63 & 1.92 & 0.21 & 2.18 & 11.0777 & 55.532 & 31 & 0 & 0.0 & 3 & 0.26 & 12.1607 & 6.116 & 31 & 0.0 & 5.50912 & 0\\
  40455 & 0.381 & 0.32 & 0.44 & 0.31 & 0.44 & 0.0459227 & 99.976 & 30 & 2 & 0.2 & 3 & -99.0 & 1.0E9 & 0.0 & -999 & -99.0 & 133.076 & 0\\
  41535 & 1.3348 & 0.34 & 1.39 & 0.33 & 1.45 & 6.26825 & 59.054 & 26 & 0 & 0.0 & 3 & 0.36 & 6.96524 & 21.341 & 31 & 0.4 & 14.71 & 0\\
\enddata
\tablenotetext{Note}{Only a portion of this table is shown here for guidance regarding its form and content. Table \ref{tab:Photoz_cat} is published in its entirety in the electronic edition of \apjs . For a detailed description see \S \ref{sec:zcat}.}
\end{deluxetable}
\end{document}